\documentclass{article}

\PassOptionsToPackage{numbers,sort&compress}{natbib}
\usepackage{resources/spml_style}

\usepackage[utf8]{inputenc}
\usepackage[T1]{fontenc}
\usepackage{url}
\usepackage{booktabs}
\usepackage{amsfonts}
\usepackage{amsmath}
\usepackage{amssymb}
\usepackage{nicefrac}
\usepackage{microtype}
\usepackage{xcolor}
\usepackage{graphicx}
\usepackage{multirow}
\usepackage{subcaption}
\usepackage{wrapfig}
\usepackage{enumitem}
\usepackage{colortbl}

\definecolor{blue(pigment)}{rgb}{0.2, 0.2, 0.6}
\definecolor{aliceblue}{rgb}{0.85, 0.9, 1.0}
\definecolor{denim}{rgb}{0.08, 0.38, 0.74}
\definecolor{lapislazuli}{rgb}{0.15, 0.38, 0.61}
\definecolor{yaleblue}{rgb}{0.06, 0.3, 0.57}
\definecolor{soupgray}{gray}{0.95}

\usepackage{xspace}
\usepackage{bm}
\usepackage{needspace}

\newcommand{\method}{\textsc{SoupFold}\xspace}

\title{Co-folding with a Soup of Representations}

\author{Hyosoon Jang}
\author{Taewon Kim}
\author{Sungsoo Ahn}
\affiliation{KAIST}
\correspondence{\email{\{hyosoon.jang,sungsoo.ahn\}@kaist.ac.kr}}

\titleabstract{Co-folding models such as AlphaFold3, Protenix, ESMFold2, and OpenDDE have advanced rapidly, yet no single model consistently performs best across all biomolecular complexes. In this paper, we show that their pair representations encode complementary information that can be transferred across models to improve structure prediction. We introduce \method{}, which combines pair representations from multiple co-folding models in a common representation space and generates structures from the combined representation. Importantly, \method{} does not retrain the co-folding models and learns only simple mappings to transfer representations across models. We evaluate \method{} on antibody-antigen, protein-protein, protein-ligand, molecular glue, GPCR, and oligomeric complex prediction using AlphaFold3, Protenix, ESMFold2, and OpenDDE. By combining representations across models, \method{} improves over individual co-folding models across the considered benchmarks.\footnote{Code: \url{https://github.com/hsjang0/soupfold}}

}

\begin{document}

\maketitle

\begin{figure}[h]
\centering
\includegraphics[width=0.95\linewidth]{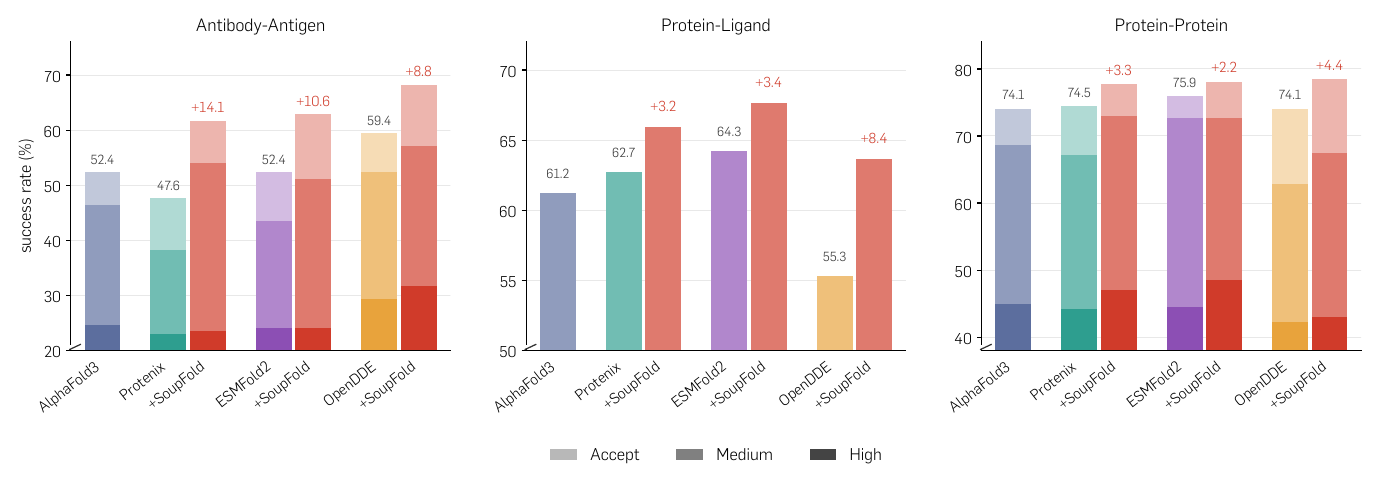}%
\caption{\textbf{Antibody-antigen, protein-ligand, and protein-protein predictions on FoldBench.} Accept, Medium, and High denote DockQ $\geq 0.23$, $\geq 0.49$, and $\geq 0.80$. Protein-ligand success denotes LRMSD $<2$~\AA\ and lDDT-PLI $>0.8$. \method{} improves performance over the baselines.}%(below) By combining complementary information across models that individually capture either the ligand pose or the TM6 active conformation, \method accurately predicts both in transducer-bound active GPCRs.
\label{fig:overview}
\end{figure}

\section{Introduction}

Co-folding models have developed rapidly in recent years. Since AlphaFold3 \citep{abramson2024af3} extended structure prediction from single proteins to general biomolecular complexes, several models have followed, including Boltz \citep{wohlwend2024boltz, passaro2025boltz2}, Chai \citep{chai2024chai1}, Protenix \citep{bytedance2025protenix}, OpenFold3 \citep{openfold3}, ESMFold2 \citep{esmfold2}, and OpenDDE \citep{project2026foldingreasoningscalingopensource}. These models share a similar pipeline: a Pairformer-based trunk constructs pair representations of a target complex, which are then used by a diffusion module to predict its structure. They mainly differ in their training recipes, including training data, objectives, and hyperparameters.

\begin{figure}[t]
\centering
\includegraphics[width=\linewidth]{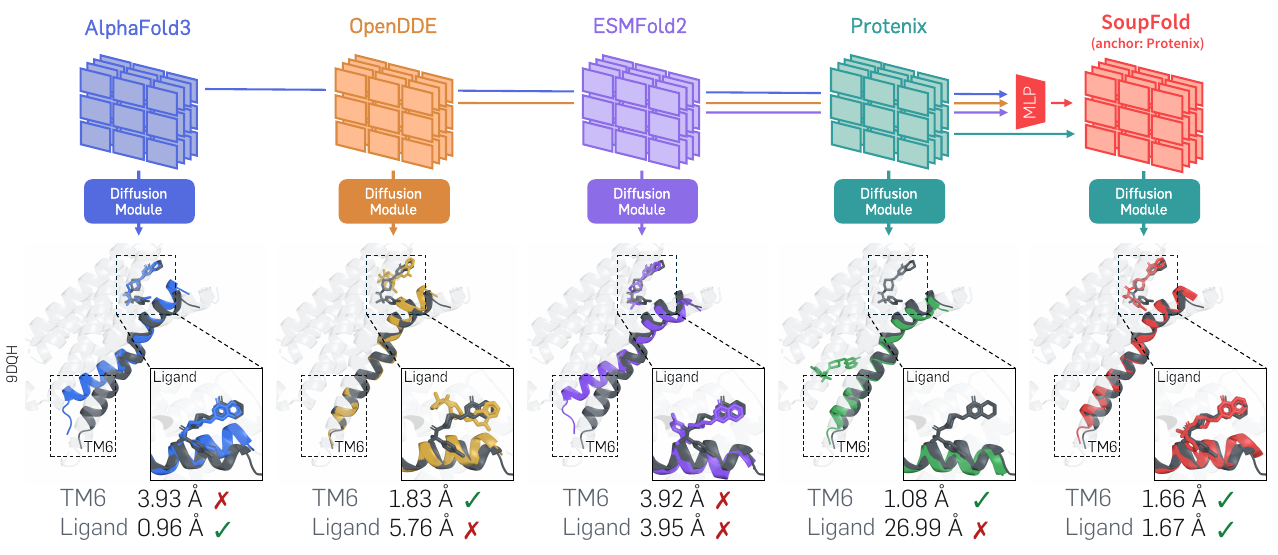} 
\vspace{-.1in}
\caption{\textbf{Examples of \method on high-order co-folding prediction.}
Accurate co-folding structure prediction of active GPCRs requires capturing both the ligand pose that induces the active state and the corresponding active-state conformation of transmembrane helix 6 (TM6). Individual co-folding models often recover only one of these features, whereas \method integrates their complementary structural information and accurately predicts both.}
\label{fig:intro}
\vspace{-.1in}
\end{figure}

With several models now addressing the same prediction tasks, ensembling offers a way to improve accuracy without training a new model from scratch. This is especially promising because no single model consistently performs best at capturing structural patterns. For example, in an active-state GPCR system (\Cref{fig:intro}), where the (1) ligand pose and (2) coupled TM6 opening need to be captured jointly, AlphaFold3 predicts (1) the correct ligand pose but (2) misplaces the TM6 opening, whereas other models show the opposite behavior. Such differences suggest that the models capture complementary structural information.

%(below) By combining complementary information across models that individually capture either the ligand pose or the TM6 active conformation, \method accurately predicts both in transducer-bound active GPCRs.

To ensemble complementary structural information across models, we note that these signals may already be encoded in the pair representations used to condition structure generation. We empirically compare representations across co-folding models and find that they are generally aligned, while their alignment weakens when the models predict different structures. These findings suggest a shared representational basis across models alongside complementary structural signals. 

Motivated by these observations, we investigate whether transferring the representations across models can improve the standalone model's predictions. We thus ask:
\begin{center}
\noindent\fcolorbox{black}{gray!15}{
\parbox{0.92\linewidth}{
\vspace{0.3em}
\centering
\textit{
Can complementary information encoded in representations be transferred across co-folding models to improve structure prediction?
}
\vspace{0.3em}
}
}
\end{center}

To this end, we propose \method{}, which ensembles representations from off-the-shelf co-folding models to generate high-quality structure predictions. Given an anchor model, lightweight transfer networks map representations from peer models into the anchor's representation space. The mapped representations are then combined with the anchor representation through a weighted average, augmenting its representation with structural signals captured by peers. The combined representation is then passed to the anchor model's diffusion module for structure generation. Importantly, \method{} does not retrain the co-folding models and only trains the lightweight transfer networks.

We evaluate \method{} using four cutting-edge co-folding models: AlphaFold3, Protenix-v1, ESMFold2, and OpenDDE. As benchmarks, we consider FoldBench \citep{foldbench2025} covering protein-protein, protein-ligand, and antibody-antigen complexes. We further consider some high-order benchmarks such as molecular glue ternary from MGBench \citep{liao2025benchmarking}, GPCR from \citet{shen2025update}, and oligomer complexes from CASP16 \citep{zhang2026assessment}. We also evaluate on two challenging benchmarks: the low-similarity subset of Runs N' Poses \citep{vskrinjar2026evaluating} for protein-ligand prediction, and ARK-AB \citep{project2026foldingreasoningscalingopensource} for antibody-antigen prediction.

Across these benchmarks, we show that
\begin{itemize}[topsep=-1.0pt,itemsep=1.0pt,leftmargin=3.5mm]
\item \method{} improves over standalone models and achieves the best or competitive performance across considered benchmarks, with larger relative gains on higher-order complexes that require multiple structural patterns to be captured jointly.
\item \method{} exploits complementary representations to improve prediction quality and can recover high-quality structures even when all standalone models fail.
\item \method{} outperforms sample-level ensembling across co-folding models and scaling standalone models with larger budgets.
\end{itemize}

\section{\method: Representation ensembling for co-folding}
\label{sec}

We introduce \method, which improves structure prediction by ensembling pair representations across multiple co-folding models. We first describe the common structure of modern co-folding models (\Cref{subsec:co-folding-rep}). We then present observations across these representations that support representation ensembling (\Cref{subsec:motivational_observations}) and describe the method (\Cref{subsec:method_overview}).

\subsection{Co-folding representations}\label{subsec:co-folding-rep}

Modern co-folding models predict the joint structure of biomolecular complexes containing proteins, nucleic acids, and small molecules. Models such as AlphaFold3 \citep{abramson2024af3}, Protenix~\citep{bytedance2025protenix}, ESMFold2~\citep{esmfold2}, and OpenDDE~\citep{project2026foldingreasoningscalingopensource} broadly decompose this task into two stages: encoding the complex into pair representations and generating structures conditioned on them.

A large trunk network first processes input features to construct pair representations, where each entry captures the structural features in biomolecules. Through recycling, the trunk iteratively refines the representations. A relatively lightweight diffusion module then generates atomic coordinates conditioned on them. At a high level, this process can be written as
\begin{equation*}
H = \operatorname{Trunk}(\mathrm{Input}),
\qquad
\mathrm{Structure} \sim \operatorname{Diffusion}(\cdot \mid H),
\end{equation*}
where $\mathrm{Input}$ includes a biomolecular token sequence of length $N$, such as atoms or residues, and multiple sequence alignments (MSAs), while $H \in \mathbb{R}^{N \times N \times D}$ denotes the pair representation encoding relationships between pairs of tokens. Our method ensembles $H$ from multiple co-folding models before generating the final structure using the $\operatorname{Diffusion}$ module. For convenience, we omit the single representations of co-folding models and do not ensemble them.

\subsection{Observations}
\label{subsec:motivational_observations}

\begin{figure}[t]
\centering\hfill
\begin{subfigure}[t]{0.555\linewidth}
\centering
\includegraphics[width=\linewidth]{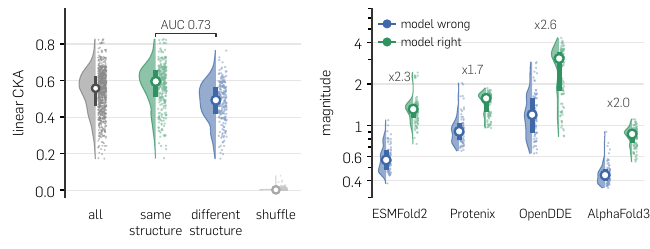}
\caption{CKA and per-model magnitude.}
\label{fig:pair_observations_cknna}
\end{subfigure}
\begin{subfigure}[t]{0.435\linewidth}
\centering
\includegraphics[width=\linewidth]{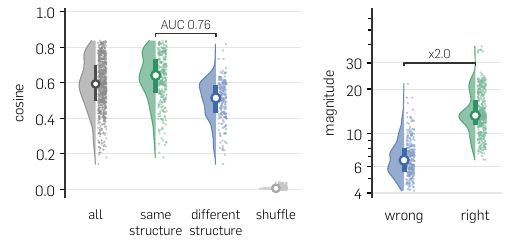}
\caption{Cosine similarity and magnitude.}
\label{fig:pair_observations_cosine}
\end{subfigure}
\caption{\textbf{Alignments and magnitudes of representations.}
(a) Representational alignment, measured by Centered Kernel Alignment (CKA) \citep{kornblith2019similarity}, across model pairs, and interface representation magnitude measured by channel-wise standard deviation. Similar and dissimilar structures denote pairs with DockQ $\geq 0.8$ and DockQ $\leq 0.23$, respectively. Alignment weakens for dissimilar structures, while unsuccessful predictions have smaller representation magnitudes. (b) Trends after mapping representations into a shared space (ESMFold2 representation space).}
\label{fig:pair_observations}
\end{figure}

We characterize pair representations across different co-folding models. We find that their representations are generally well aligned, but alignment weakens when they predict different structures, suggesting that they share a representational basis while retaining potentially complementary structural signals. Meanwhile, unsuccessful representations tend to have smaller magnitudes, reducing the influence of potentially conflicting signals when ensembled.

\paragraph{Selectively aligned representations support ensembling.}
Pair representations from different co-folding models are generally well aligned. Their alignment is stronger when the models predict similar structures and weaker when their predictions differ (\Cref{fig:pair_observations_cknna}, left). This suggests that the representations share common structure while capturing complementary structural information when predictions diverge, providing a basis for representation ensembling.

\paragraph{Unsuccessful predictions correlate with small magnitude of representations.}
When grouped by prediction quality, representations associated with unsuccessful predictions tend to have smaller magnitudes, quantified by the channel-wise standard deviation at the interface (\Cref{fig:pair_observations_cknna}, right). These smaller magnitudes correspond to less differentiated patterns across channels, suggesting weaker structural signals rather than strong conflicting ones. Their weaker signals are therefore expected to have less influence than stronger signals from successful predictions when combined.

%\paragraph{Mapping preserves the properties that support ensembling.} After mapping into another model's representation space, both trends remain (\Cref{fig:pair_observations_cosine}). Mapped representations are more aligned when the models predict similar structures and less aligned when their predictions differ, while unsuccessful predictions retain smaller magnitudes. Thus, mapping representations into a shared space allows them to be ensembled and directly processed by the existing diffusion module.

\paragraph{Representations can be mapped and aligned in a common space.}
We find that representations from different co-folding models can be meaningfully mapped from one model to another using a simple multi-layer perceptrons (MLPs). The mapped representations retain the patterns observed above: they are more aligned when the models predict similar structures, less aligned when their predictions differ, and smaller in magnitude for unsuccessful predictions (\Cref{fig:pair_observations_cosine}).

\subsection{Method}
\label{subsec:method_overview}

From the observations in \Cref{subsec:motivational_observations}, we ensemble representations across multiple co-folding models. Given a set of co-folding models $\mathcal{M}$, we choose one model $m^\star\in\mathcal{M}$ as the anchor model, whose diffusion module generates the structure, and let the remaining models $\mathcal{M}\setminus{m^\star}$ be peers that provide additional representations. For each peer $m\in\mathcal{M}\setminus\{{m^\star}\}$, we train a transfer network $f_{m\to m^\star}$ that maps its representation into the anchor representation space for ensembling.

At inference time, we combine the representations by a weighted average in representation space of the anchor model. Each peer model's representation $H^{(m)}$ is mapped into the anchor model's representation space through a peer-to-anchor map $f_{m\to m^\star}$, and the peer representations are averaged with the anchor representation:
\begin{align*}
H=
\frac{
w_{m^\star} H^{(m^\star)}+\sum_{m\in\mathcal{M}\setminus \{m^\star\}} w_m f_{m\to m^\star}(H^{(m)})
}{
w_{m^\star}+\sum_{m\in\mathcal{M}\setminus \{m^\star\}} w_m
}
\end{align*}
where $w_{m^\star}$ and $w_m$ are the coefficients for the anchor model and each peer, respectively. The updated representation $H$ is then passed to the anchor model's diffusion module to predict the structure. In all experiments, we use $w_{m^{\star}}=1$ and $w_m=1$ for all peers by default.

\paragraph{Training transfer networks.} We train a transfer network $f_{m\rightarrow m^\star}$ for each directed pair of co-folding models. Each map is a one-layer MLP of width $1024$, applied independently to each pair entry. Given representations from two co-folding models for the same input, we train each transfer network with the following mean-squared error loss:
\begin{equation*}
\mathcal{L}(f_{m\rightarrow m^\star})
=
\mathbb{E}_{(H^{(m)},H^{(m^\star)})\in\mathcal{D}}
\mathbb{E}_{(i,j)}
\left[
\left\|
f_{m\rightarrow m^\star}(H^{(m)}_{ij})
-
H^{(m^\star)}_{ij}
\right\|_2^2
\right].
\end{equation*}
Here, $H^{(m)}_{ij}$ denotes the pair representation of model $m$ for token pair $(i,j)$. Note that we apply channel-wise normalization for training stability. The representation training set $\mathcal{D}$ is constructed from RCSB entries released before September 30, 2021, following the conventional training cutoff used by co-folding models and remaining temporally separated from most evaluation benchmarks. We provide additional details in \Cref{sec:soupfold_appendix}.%, including the training of the transfer networks.

% ============================================================
% Shared command
% ============================================================

\providecommand{\mainnewscore}[3]{%
  \shortstack[c]{#1\\{\scriptsize #2\,$\pm$\,#3}}%
}

\providecommand{\mainnewscoregain}[4]{%
  \shortstack[c]{%
    #1\,{\tiny $\uparrow$\,#4}\\%
    {\scriptsize #2\,$\pm$\,#3}%
  }%
}

\providecommand{\mainnewscoredrop}[4]{%
  \shortstack[c]{%
    #1\,{\tiny $\downarrow$\,#4}\\%
    {\scriptsize #2\,$\pm$\,#3}%
  }%
}

\providecommand{\methodrow}{\raisebox{1.2ex}[0pt][0pt]{+\method}}

% ============================================================
% Table 1: FoldBench
% ============================================================

\begin{table*}[t]

\centering

\providecommand{\maintableformat}{\fontsize{4.0}{10}\selectfont\setlength{\tabcolsep}{3.5pt}}

\setlength{\tabcolsep}{3.5pt}

\definecolor{soupgray}{gray}{0.95}

\providecommand{\method}{SoupFold}

\caption{%
\textbf{Co-folding prediction on FoldBench.}
For antibody-antigen (170 interfaces) and protein-protein (274 interfaces) prediction, Accept, Medium, and High denote DockQ $\geq 0.23$, $\geq 0.49$, and $\geq 0.80$. Protein-ligand success (526 interfaces) requires LRMSD $<2$\AA\ and lDDT-PLI $>0.8$. The top line in each cell reports the best-confidence-of-25 result (5 samples per seed, 5 seeds) with gains of \method{} over the anchor denoted with $\uparrow$. The bottom line reports the mean $\pm$ standard deviation over the 25 samples. \method{} consistently improves the anchor model, and achieves the best performance. \textbf{Bold} indicates the better result within each anchor.
}
%\vspace{-.05in}
\label{tab:soupfold_foldbench}

\begin{tabular}{l ccc c ccc}

\toprule

& \multicolumn{3}{c}{AbAg}
& \multicolumn{1}{c}{PL}
& \multicolumn{3}{c}{PP}
\tabularnewline

\cmidrule(lr){2-4}
\cmidrule(lr){5-5}
\cmidrule(lr){6-8}

Model
& Accept & Medium & High
& Success
& Accept & Medium & High
\tabularnewline

\midrule

AlphaFold3
& \mainnewscore{52.35}{43.25}{2.14}
& \mainnewscore{46.47}{35.95}{1.73}
& \mainnewscore{24.71}{18.07}{1.75}
& \mainnewscore{61.22}{62.16}{0.68}
& \mainnewscore{74.09}{74.09}{0.73}
& \mainnewscore{68.61}{68.47}{0.20}
& \mainnewscore{44.89}{44.45}{1.35}
\tabularnewline

\midrule

Protenix
& \mainnewscore{47.65}{42.73}{2.57}
& \mainnewscore{38.24}{34.85}{1.55}
& \mainnewscore{22.94}{18.54}{2.24}
& \mainnewscore{62.74}{61.54}{0.77}
& \mainnewscore{74.45}{72.34}{1.12}
& \mainnewscore{67.15}{64.34}{1.57}
& \mainnewscore{44.16}{42.18}{1.53}
\tabularnewline

\rowcolor{soupgray}
\methodrow
& \mainnewscoregain{\textbf{61.76}}{55.53}{2.79}{14.11}
& \mainnewscoregain{\textbf{54.12}}{45.60}{2.35}{15.88}
& \mainnewscoregain{\textbf{23.53}}{20.42}{1.82}{0.59}
& \mainnewscoregain{\textbf{65.97}}{65.16}{1.00}{3.23}
& \mainnewscoregain{\textbf{77.74}}{75.69}{1.37}{3.29}
& \mainnewscoregain{\textbf{72.99}}{70.25}{1.26}{5.84}
& \mainnewscoregain{\textbf{47.08}}{44.35}{1.67}{2.92}
\tabularnewline

\midrule

ESMFold2
& \mainnewscore{52.35}{46.66}{3.10}
& \mainnewscore{43.53}{37.48}{2.68}
& \mainnewscore{\textbf{24.12}}{18.59}{1.58}
& \mainnewscore{64.26}{62.94}{0.66}
& \mainnewscore{75.91}{73.14}{1.16}
& \mainnewscore{\textbf{72.63}}{67.68}{1.12}
& \mainnewscore{44.53}{38.73}{1.75}
\tabularnewline

\rowcolor{soupgray}
\methodrow
& \mainnewscoregain{\textbf{62.94}}{52.96}{2.47}{10.59}
& \mainnewscoregain{\textbf{51.18}}{43.58}{1.71}{7.64}
& \mainnewscoregain{\textbf{24.12}}{20.12}{1.70}{0.00}
& \mainnewscoregain{\textbf{67.68}}{65.40}{0.92}{3.42}
& \mainnewscoregain{\textbf{78.10}}{75.33}{0.94}{2.19}
& \mainnewscoregain{\textbf{72.63}}{70.22}{0.95}{0.00}
& \mainnewscoregain{\textbf{48.54}}{45.07}{1.79}{4.01}
\tabularnewline

\midrule

OpenDDE
& \mainnewscore{59.41}{56.16}{1.91}
& \mainnewscore{52.35}{47.76}{1.57}
& \mainnewscore{29.41}{25.48}{1.84}
& \mainnewscore{55.32}{54.43}{0.71}
& \mainnewscore{74.09}{72.74}{0.64}
& \mainnewscore{62.77}{61.80}{0.58}
& \mainnewscore{42.34}{39.56}{0.99}
\tabularnewline

\rowcolor{soupgray}
\methodrow
& \mainnewscoregain{\textbf{68.24}}{61.51}{1.84}{8.83}
& \mainnewscoregain{\textbf{57.06}}{52.05}{1.26}{4.71}
& \mainnewscoregain{\textbf{31.76}}{28.14}{1.38}{2.35}
& \mainnewscoregain{\textbf{63.69}}{62.71}{0.81}{8.37}
& \mainnewscoregain{\textbf{78.47}}{77.37}{0.82}{4.38}
& \mainnewscoregain{\textbf{67.52}}{66.88}{0.66}{4.75}
& \mainnewscoregain{\textbf{43.07}}{41.58}{1.20}{0.73}
\tabularnewline
\bottomrule
\end{tabular}
\end{table*}

\begin{figure}[t]
\centering
\centering\includegraphics[width=\linewidth]{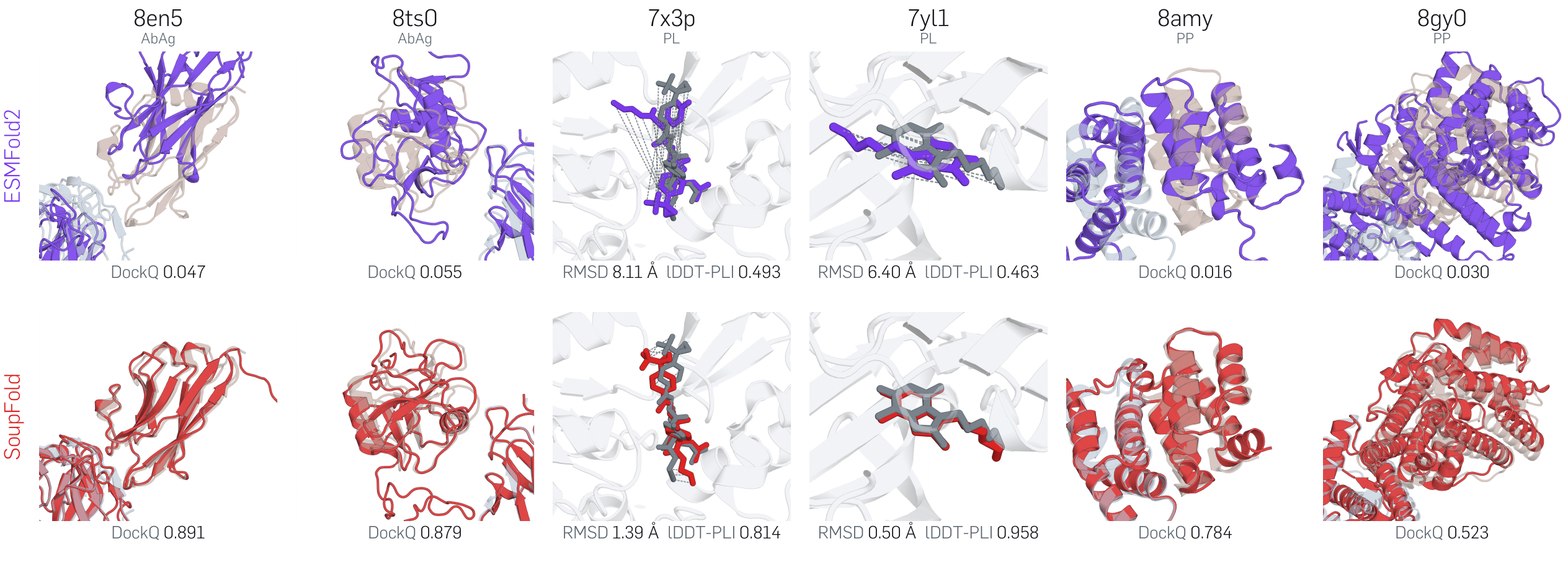}
%\vspace{-.1in}
\caption{\textbf{\method recovers failure predictions using peer representations.} Examples of antibody-antigen, protein-protein, and protein-ligand interfaces whose predictions are rescued by \method using other peer representations.}
%\vspace{-.1in}
\label{fig:rescue_samples}
\end{figure}

\section{Results}
\label{sec:exp}

We evaluate whether \method can improve protein co-folding structure prediction by combining representations from multiple co-folding models. We consider AlphaFold3 \citep{abramson2024af3}, Protenix-v1 \citep{protenix-v1}, ESMFold2~\citep{esmfold2}, and OpenDDE \citep{project2026foldingreasoningscalingopensource} on conventional antibody-antigen, protein-ligand, and protein-protein benchmarks, as well as higher-order complex benchmarks including molecular glue, GPCR, and oligomer complexes. We use Protenix-v1, ESMFold2, and OpenDDE as anchor models in turn, ensembling their representations with those of the other three peer models and generating structures using their diffusion modules. We do not use AlphaFold3 as an anchor model due to difficulties injecting representations into its JAX-compiled graph, but include it as a peer.

\paragraph{Configuration.} All models use unpaired and paired MSAs, $10$ recycling steps, and their default diffusion steps. We use templates for AlphaFold3, Protenix, and OpenDDE, while ESMFold2 does not support the use of templates. Templates are prepared from PDBe structures released on or before September 30, 2021. The details are provided in \Cref{appx:configuration}. 

\paragraph{Benchmark.}
On FoldBench \citep{foldbench2025}, we report DockQ-based success for antibody-antigen and protein-protein complexes, and LRMSD and lDDT-PLI-based success for protein-ligand complexes. We further evaluate higher-order complexes using MGBench molecular glue benchmark \citep{liao2025benchmarking}, GPCR benchmark~\citep{shen2025update}, and CASP16 oligomer benchmark \citep{zhang2026assessment}. MGBench evaluates protein-protein and protein-ligand success, while GPCR evaluates ligand-pose success and TM6 prediction. For CASP16 oligomers, we report TM-score, QSbest, DockQ, lDDT, ICS, and IPS. We additionally evaluate ARK-AB \citep{project2026foldingreasoningscalingopensource} and the low-similarity subset of Runs N' Poses \citep{vskrinjar2026evaluating} for challenging antibody-antigen and protein-ligand prediction, respectively. Detailed settings are provided in \Cref{appx:benchmark}.

%For oligomers, we consider complexes with three or more chains and report TM-score, QSbest, and DockQ. Additionally, we evaluate challenging antibody-antigen and protein-ligand benchmarks using ARK-AB and the low-similarity subset of Runs N' Poses, respectively. Detailed benchmark settings are provided in \Cref{appx:appendix_benchmark}.

\providecommand{\mainnewscoregainoff}[4]{%
  \shortstack[c]{%
    #1 \\ % %#1\,{\tiny $\uparrow$\,#4}\\%
    {\scriptsize #2\,$\pm$\,#3}%
  }%
}

\begin{table*}[t]

\providecommand{\maintableformat}{\fontsize{4.0}{10}\selectfont\setlength{\tabcolsep}{2.0pt}}

\centering
\setlength{\tabcolsep}{2.0pt}

\caption{%
\textbf{Co-folding prediction on molecular-glue, GPCR, and oligomer benchmarks.}
For molecular-glue ($88$ complexes) prediction, PLP (ternary success) requires PP success and PL success simultaneously.
For GPCR ($69$ complexes) prediction, TM6-PL success requires the RMSD of the TM6 tip to be $<3.0$~\AA\ and PL success simultaneously. 
For oligomer ($18$ complexes) prediction, we report TM-score, QSbest, and DockQ.
We report the best-confidence result (5 samples per seed, 5 seeds), and the mean $\pm$ standard deviation.
See \Cref{appx:mgbench_results,appx:gpcr_results,appx:casp16_results} for full tables.
}
\label{tab:soupfold_large}

\begin{tabular}{l ccc ccc ccc}
\toprule
& \multicolumn{3}{c}{Molecular Glue}
& \multicolumn{3}{c}{GPCR}
& \multicolumn{3}{c}{Oligomer}
\tabularnewline

\cmidrule(lr){2-4}
\cmidrule(lr){5-7}
\cmidrule(lr){8-10}

Model
& PLP & PP & PL
& TM6-PL & TM6 & PL
& TM & QSbest & DockQ
\tabularnewline

\midrule

AlphaFold3
& \mainnewscore{31.82}{33.9}{1.1}
& \mainnewscore{47.73}{49.6}{2.7}
& \mainnewscore{32.95}{34.1}{1.0}
& \mainnewscore{49.28}{48.5}{2.6}
& \mainnewscore{1.24}{1.22}{.01}
& \mainnewscore{53.62}{50.9}{2.3}
& \mainnewscore{0.838}{0.802}{.017}
& \mainnewscore{0.637}{0.642}{.016}
& \mainnewscore{0.511}{0.478}{.023}
\tabularnewline

\midrule

Protenix
& \mainnewscore{28.41}{31.2}{2.4}
& \mainnewscore{50.00}{49.5}{1.3}
& \mainnewscore{28.41}{31.5}{2.5}
& \mainnewscore{49.28}{50.6}{2.7}
& \mainnewscore{1.32}{1.34}{.04}
& \mainnewscore{50.72}{53.7}{2.5}
& \mainnewscore{0.826}{0.760}{.026}
& \mainnewscore{0.655}{0.616}{.024}
& \mainnewscore{0.488}{0.443}{.019}
\tabularnewline

\rowcolor{soupgray}
+\method
& \mainnewscoregainoff{\textbf{36.36}}{35.9}{1.2}{7.95}
& \mainnewscoregainoff{\textbf{54.55}}{51.9}{1.3}{4.55}
& \mainnewscoregainoff{\textbf{36.36}}{36.1}{1.2}{7.95}
& \mainnewscoregainoff{\textbf{56.52}}{51.0}{3.3}{7.25}
& \mainnewscoregainoff{\textbf{1.19}}{1.20}{.02}{.13}
& \mainnewscoregainoff{\textbf{57.97}}{53.3}{2.9}{7.25}
& \mainnewscoregainoff{\textbf{0.832}}{0.815}{.026}{.006}
& \mainnewscoregainoff{\textbf{0.697}}{0.657}{.025}{.042}
& \mainnewscoregainoff{\textbf{0.551}}{0.485}{.026}{.063}
\tabularnewline

\midrule

ESMFold2
& \mainnewscore{30.68}{26.6}{2.2}
& \mainnewscore{\textbf{53.41}}{48.6}{1.4}
& \mainnewscore{31.82}{26.6}{2.2}
& \mainnewscore{43.48}{42.4}{2.6}
& \mainnewscore{1.33}{1.36}{.03}
& \mainnewscore{47.83}{45.9}{2.2}
& \mainnewscore{0.752}{0.693}{.033}
& \mainnewscore{0.605}{0.570}{.031}
& \mainnewscore{0.450}{0.405}{.027}
\tabularnewline

\rowcolor{soupgray}
+\method
& \mainnewscoregainoff{\textbf{39.77}}{34.1}{2.7}{9.09}
& \mainnewscoregainoff{52.27}{51.4}{1.7}{1.14}
& \mainnewscoregainoff{\textbf{39.77}}{34.2}{2.8}{7.95}
& \mainnewscoregainoff{\textbf{50.72}}{48.5}{2.1}{7.25}
& \mainnewscoregainoff{\textbf{1.19}}{1.19}{.02}{.15}
& \mainnewscoregainoff{\textbf{53.62}}{50.2}{2.4}{5.80}
& \mainnewscoregainoff{\textbf{0.818}}{0.757}{.026}{.066}
& \mainnewscoregainoff{\textbf{0.659}}{0.603}{.031}{.054}
& \mainnewscoregainoff{\textbf{0.489}}{0.446}{.020}{.039}
\tabularnewline

\midrule

OpenDDE
& \mainnewscore{\textbf{31.82}}{31.1}{1.6}
& \mainnewscore{50.00}{57.6}{3.1}
& \mainnewscore{\textbf{31.82}}{31.1}{1.6}
& \mainnewscore{46.38}{42.8}{2.8}
& \mainnewscore{1.39}{1.35}{.03}
& \mainnewscore{47.83}{46.5}{2.2}
& \mainnewscore{0.651}{0.658}{.023}
& \mainnewscore{0.506}{0.499}{.013}
& \mainnewscore{0.454}{0.447}{.012}
\tabularnewline

\rowcolor{soupgray}
+\method
& \mainnewscoregainoff{27.27}{27.4}{1.9}{4.55}
& \mainnewscoregainoff{\textbf{52.27}}{55.2}{2.5}{2.27}
& \mainnewscoregainoff{27.27}{27.4}{1.9}{4.55}
& \mainnewscoregainoff{\textbf{49.28}}{49.5}{2.5}{2.90}
& \mainnewscoregainoff{\textbf{1.18}}{1.21}{.03}{.20}
& \mainnewscoregainoff{\textbf{50.72}}{51.3}{2.2}{2.90}
& \mainnewscoregainoff{\textbf{0.721}}{0.687}{.018}{.070}
& \mainnewscoregainoff{\textbf{0.569}}{0.529}{.019}{.063}
& \mainnewscoregainoff{\textbf{0.530}}{0.475}{.020}{.076}
\tabularnewline

\bottomrule
\end{tabular}
\vspace{-.15in}
\end{table*}

%\input{resources/Table_main_additional}

%\subsection{\method Improves Prediction Using Representations from Other Models}
\subsection{Main results}
\label{subsec:representation_transfer}

\begin{wraptable}{r}{0.35\textwidth}

\vspace{-0.15in}

\centering

\fontsize{9.0}{10}\selectfont

\setlength{\tabcolsep}{5.0pt}

\renewcommand{\arraystretch}{1.0}

\definecolor{soupgray}{gray}{0.95}

\providecommand{\method}{SoupFold}

\caption{\textbf{Additional challenging benchmarks.} See full results in \Cref{appx:rnp_results,appx:ark_results}.}
\label{tab:rnp_ark}
\vspace{-0.1in}
\begin{tabular}{l cc}
\toprule
Model
& ARK
& RnP$^\ast$
\tabularnewline
\midrule
AlphaFold3
& 47.66
& 28.89
\tabularnewline
\midrule
Protenix
& 45.94
& 21.11
\tabularnewline
\rowcolor{soupgray}
+\method
& \textbf{56.08}
& \textbf{25.56}
\tabularnewline
\midrule
ESMFold2
& 48.68
& 27.78
\tabularnewline
\rowcolor{soupgray}
+\method
& \textbf{60.53}
& \textbf{28.89}
\tabularnewline
\midrule
OpenDDE
& 58.92
& 15.56
\tabularnewline
\rowcolor{soupgray}
+\method
& \textbf{64.35}
& \textbf{24.44}
\tabularnewline
\bottomrule
\end{tabular}
\vspace{-0.1in}
\end{wraptable}

We evaluate whether representations from other models can improve the performance of an anchor model. In \Cref{tab:soupfold_foldbench}, \method{} consistently improves antibody-antigen, protein-ligand, and protein-protein prediction. For each anchor model, \method{} outperforms or is competitive with the strongest standalone model, with large gains on antibody-antigen prediction. These results suggest that independently trained models provide complementary representations that improve predictions.

\begin{figure*}
  \begin{subfigure}[t]{0.65\textwidth}
    \includegraphics[width=\linewidth]{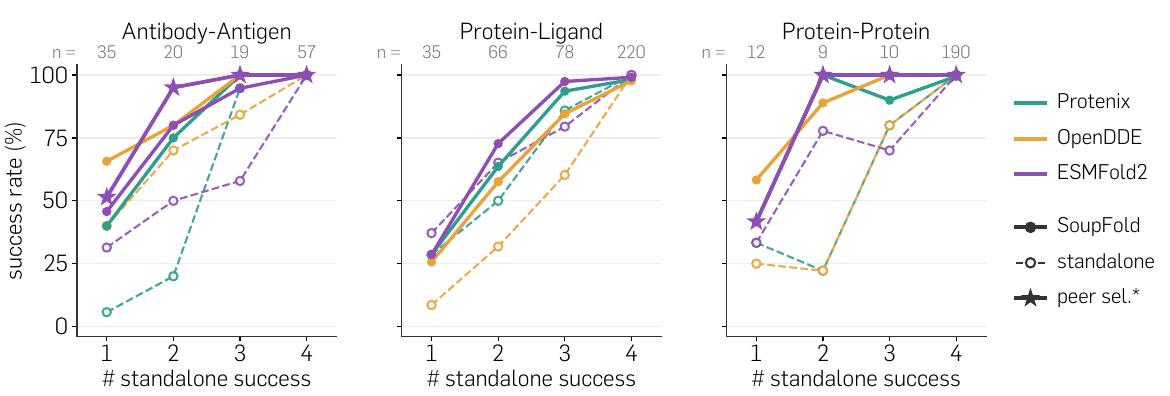}
    \caption{Success by the number of successful standalone models.}\label{fig:succ_vs_soup}
  \end{subfigure}
  \begin{subfigure}[t]{0.34\textwidth}
    \includegraphics[width=\linewidth]{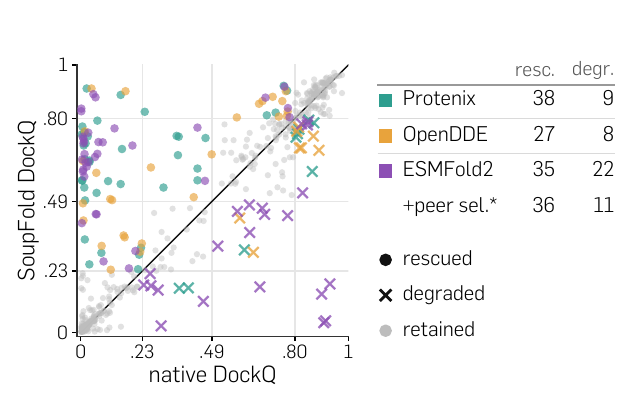}
    \caption{Sample-wise DockQ changes.}\label{fig:dockq_change}
  \end{subfigure}\hfill
   \vspace{-.1in}
  \caption{\textbf{Gains from \method.} (a) \method improves even when only two models succeed, consistent with failed representations having less influence (\Cref{subsec:motivational_observations}). (b) \method improves more antibody-antigen targets than it degrades. Peer sel.$^{\star}$ denotes the variant in \Cref{tab:metric_ablation}.}
 \vspace{-.2in}
\end{figure*}

The performance gains are also notable on the higher-order complexes in \Cref{tab:soupfold_large}. For molecular glue-induced ternary complexes, \method{} substantially improves joint success requiring both protein-protein and protein-ligand predictions to succeed. Similar gains are observed for GPCR complexes, where joint success requires both accurate ligand-pose and TM6-branch prediction. For oligomer complexes, Protenix+\method{} matches AlphaFold3 in TM-score while improving QSbest and DockQ, indicating complementary gains in global and interface quality. Additionally, \method{} consistently improves over the anchor models on ARK-AB and the low-similarity subset of Runs N' Poses (\Cref{tab:rnp_ark}). On Runs N' Poses, \method matches AlphaFold3 in success rate but further improves RMSD (\Cref{appx:rnp_results}).

\begin{figure}[t]
\setcounter{figure}{7}
\centering
\centering\includegraphics[width=\linewidth]{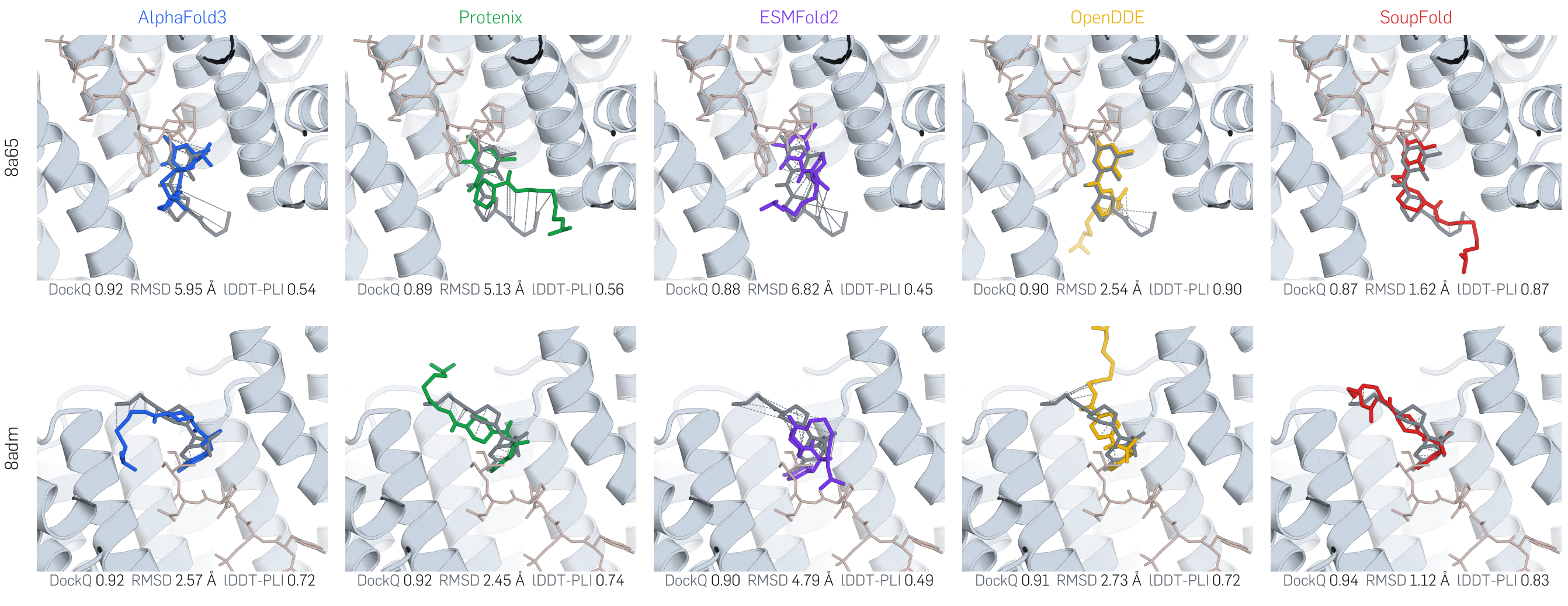}
\caption{\textbf{\method succeeds even when all standalone co-folding models strictly fail on molecular glue ternary complexes.} Molecular glue examples where \method succeeds with its highest-confidence prediction, while all standalone models fail. For 8a65, \method succeeds with ESMFold2 and OpenDDE, and for 8adm, with all three anchors. Notably, the standalone models still fail even under oracle selection over one hundred generated samples.}
\label{fig:molecular_glue}
\end{figure}

\begin{figure}[t]
\centering
\centering\includegraphics[width=\linewidth]{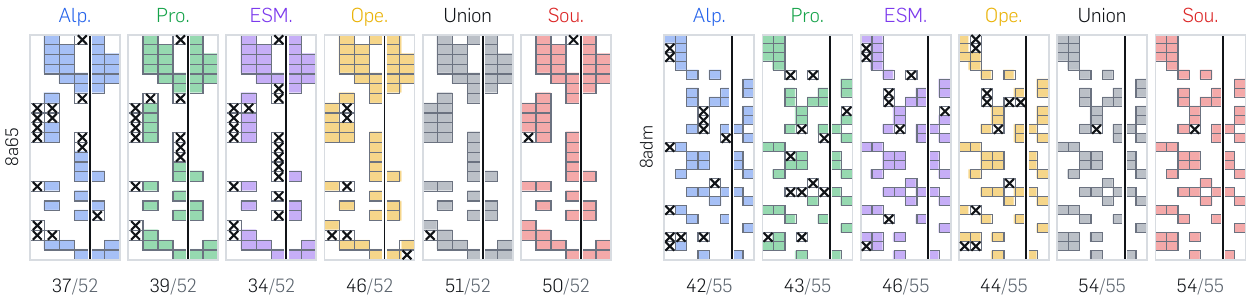}
\vspace{-.12in}
\caption{%
\textbf{\method recovers complete interfaces (\Cref{fig:molecular_glue}) by combining complementary information scattered across representations.}
We illustrate residue-atom contacts for examples in \Cref{fig:molecular_glue}.
A contact is filled when it is recovered in at least 5 of 25 samples, while $\boxtimes$ marks a missed contact.
Each standalone model captures only a subset of the contacts, while their union recovers nearly all ground-truth contacts.
This suggests that \method{} combines complementary contacts across model representations to recover the high-quality structure.
}
\label{fig:molecular_glue_explain}
\end{figure}

\subsection{Performance gains are associated with complementary representations}
\label{subsec:complementarity}
\label{sec:exp_3}

\begin{wrapfigure}{r}{0.35\textwidth}
\setcounter{figure}{5}
\vspace{-0.13in}
\centering
\includegraphics[width=0.9\linewidth]{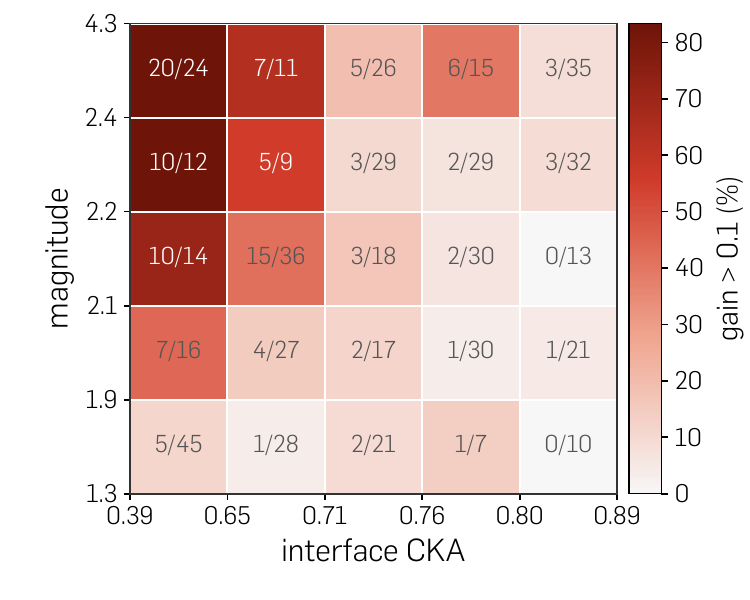} \vspace{-.12in}
\caption{\textbf{CKA and magnitude vs. DockQ gain.} Dissimilar high-magnitude representation benefits.}
\label{fig:representation_complementarity}
\vspace{-0.16in}
\end{wrapfigure}

We further examine when \method{} improves prediction quality. As shown in \Cref{fig:succ_vs_soup}, \method{} improves over the standalone models even when only two of four models succeed, consistent with unsuccessful representations having less influence on the ensemble (\Cref{subsec:motivational_observations}). \Cref{fig:dockq_change} further shows that, although most predictions retain their original interface quality, improvements are more frequent than degradations. However, the benefit varies across targets.

We hypothesize that larger gains stem from stronger complementary signals, giving \method{} more useful information to exploit. We compare DockQ gains with representation similarity between anchor and peers, shown  in \Cref{fig:representation_complementarity}. Here, the similarity is calculated via CKA. As hypothesized, we find gains to be larger when peer representations are less aligned with the anchor and have larger magnitudes, suggesting that stronger complementary representations can provide greater benefit.

%\paragraph{\method Recovers Predictions Even When All Individual Models Fail.} More interestingly, we also find cases where all individual co-folding models fail, but \method successfully predicts the structure. We qualitatively illustrate these cases in \Cref{fig:rescue_challenging}. Here, the improvement cannot be explained by transferring high-quality representation of a peer that already predicts the correct structure. Instead, combining representations from multiple models can recover useful information that is not sufficient for successful prediction in any individual model. These results provide further evidence that co-folding models encode complementary information that can be combined to produce acceptable structure predictions.

\begin{wrapfigure}{r}{0.35\textwidth}
\centering
\vspace{.15in}
\includegraphics[width=\linewidth]{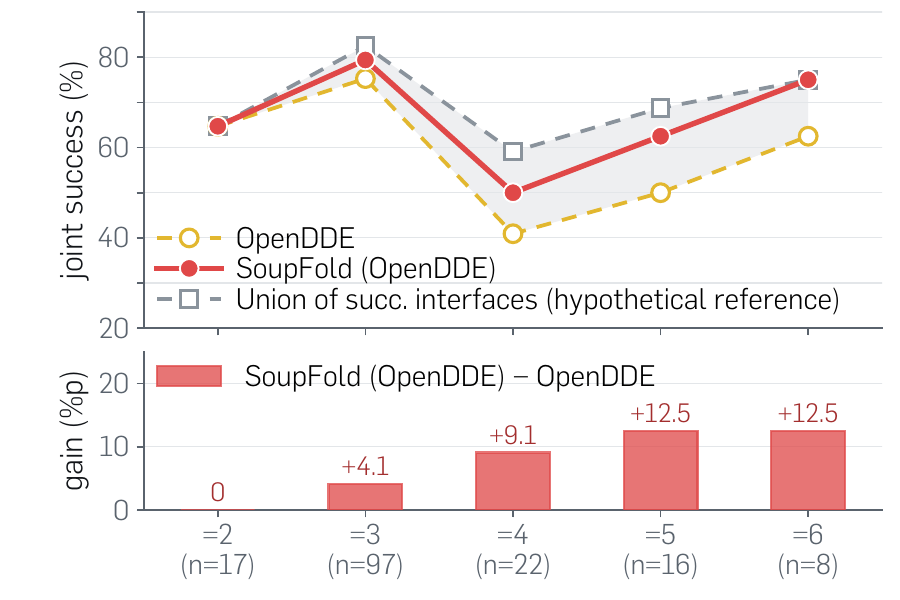}
\vspace{-.2in}
\caption{\textbf{Entry-level success for multi-interface entries.} Gains increase with the number of chains.}
\label{fig:qunati_compli}
\vspace{-.1in}
\end{wrapfigure}

\paragraph{\method combines low-quality standalone representations into high-quality predictions.}

\method is effective for higher-order complexes (\Cref{tab:soupfold_large}), which require multiple structural patterns to be captured jointly and can benefit from combining partial information across standalone models. For example, in the GPCR example in \Cref{fig:intro}, standalone models correctly capture either the ligand pose or the TM6 opening, but not both, while \method{} recovers both by integrating complementary information. 

We examine this benefit by extending the standard interface-level evaluation of ARK-AB to the entry level, where success requires all interfaces in an entry to have DockQ $>0.23$. As entries contain more chains and interfaces, individual models may capture only subsets of them, while \method{} can combine complementary information across models to recover the full interfaces. In \Cref{fig:qunati_compli}, the gain from \method{} increases with the number of chains, while its performance approaches a hypothetical upper bound given by the union of successful interfaces from all standalone models.

More interestingly, complementarity also appears at a finer interface level. As shown in \Cref{fig:molecular_glue,fig:molecular_glue_explain}, each standalone model recovers only a subset of native contacts, while their union covers nearly all contacts required for the correct structure. Thus, even when no individual model produces a successful prediction, their representations contain complementary interface signals that \method{} integrates to recover a high-quality structure.

\begin{table*}[t]
\centering
\small
\setlength{\tabcolsep}{4.5pt}
\renewcommand{\arraystretch}{1.0}
\setlength{\aboverulesep}{1.5pt}
\setlength{\belowrulesep}{2pt}
\captionsetup{position=top,skip=6pt}
\newcommand{\metricgain}[1]{{\scriptsize $\uparrow$\,#1}}
\newcommand{\metricloss}[1]{{\scriptsize $\downarrow$\,#1}}
\caption{\textbf{Teacher selection by representation complementarity.}
ESMFold2 is the anchor model. 
Peers are ranked by increasing CKA to the anchor representation, measured after mapping. 
Top-1 and Top-2 use the first one and two teachers, respectively.
Equal uses equal weights. Weighted scales each by its interface magnitude.
Accept, medium, and high rates follow the DockQ thresholds in \Cref{tab:soupfold_foldbench}. Gains over the anchor are denoted by $\uparrow$.}
\label{tab:metric_ablation}
\begin{tabular}{ll ccc ccc}
\toprule
& & \multicolumn{3}{c}{Antibody-Antigen}
& \multicolumn{3}{c}{Protein-Protein} \tabularnewline
\cmidrule(lr){3-5}
\cmidrule(lr){6-8}
Criterion & Selection
& Accept & Medium & High
& Accept & Medium & High \tabularnewline
\midrule
\multicolumn{2}{l}{ESMFold2 (No Peer)}
& 52.35 & 43.53 & 24.12
& 75.91 & 72.63 & 44.53 \tabularnewline
\cmidrule(l){2-8}
\multirow{4}{*}{Equal}
& All
& 62.94 \metricgain{10.6}
& 51.18 \metricgain{7.6}
& 24.12 \metricgain{0.0}
& 78.10 \metricgain{2.2}
& 72.63 \metricgain{0.0}
& 48.54 \metricgain{4.0} \tabularnewline
& Top-1
& 65.29 \metricgain{12.9}
& 55.88 \metricgain{12.3}
& 25.88 \metricgain{1.8}
& 78.83 \metricgain{2.9}
& 73.36 \metricgain{0.7}
& 45.26 \metricgain{0.7} \tabularnewline
& Top-2
& 62.94 \metricgain{10.6}
& 52.94 \metricgain{9.4}
& 26.47 \metricgain{2.3}
& 78.83 \metricgain{2.9}
& 72.63 \metricgain{0.00}
& 47.08 \metricgain{2.5} \tabularnewline
%& Bottom-1
%& 59.41 \metricgain{7.1}
%& 47.06 \metricgain{3.5}
%& 25.29 \metricgain{1.2}
%& 76.28 \metricgain{0.4}
%& 71.53 \metricloss{1.1}
%& 46.35 \metricgain{1.8} \tabularnewline
\cmidrule(l){2-8}
\multirow{4}{*}{Weighted}
& All
& 64.12 \metricgain{11.8}
& 54.12 \metricgain{10.6}
& 25.88 \metricgain{1.8}
& 77.74 \metricgain{1.8}
& 72.63 \metricgain{0.0}
& 48.91 \metricgain{4.4} \tabularnewline
& Top-1
& 66.47 \metricgain{14.1}
& 55.88 \metricgain{12.3}
& 26.47 \metricgain{2.3}
& 79.20 \metricgain{3.3}
& 74.09 \metricgain{1.5}
& 43.80 \metricloss{0.7} \tabularnewline
& Top-2
& 62.94 \metricgain{10.6}
& 52.94 \metricgain{9.4}
& 26.47 \metricgain{2.3}
& 77.74 \metricgain{1.8}
& 72.26 \metricloss{0.4}
& 47.08 \metricgain{2.5} \tabularnewline
%& Bottom-1
%& 56.47 \metricgain{4.1}
%& 47.06 \metricgain{3.5}
%& 25.29 \metricgain{1.2}
%& 76.28 \metricgain{0.4}
%& 72.26 \metricloss{0.4}
%& 45.99 \metricgain{1.5} \tabularnewline
\bottomrule
\end{tabular}
\vspace{-.03in}
\end{table*}

\begin{table*}[t]
\centering
\small
\setlength{\tabcolsep}{4.4pt}
\renewcommand{\arraystretch}{1.0}
\setlength{\aboverulesep}{1.5pt}
\setlength{\belowrulesep}{2pt}
\captionsetup{position=top,skip=6pt}

\caption{\textbf{\method{} vs. best-confidence selection from four pooled standalone models.} We consider FoldBench, MGBench ternary complexes, CASP16 oligomers, ARK, and the low-similarity subset of Runs N' Poses. \method{} remains competitive with or outperforms.}\label{tab:pooled}
\label{tab:pooled_samples}

\begin{tabular}{l cccc cccc}
\toprule
Method
& AbAg 
& PL
& PP
& MG
& GPCR
& Oligo.
& AbAg$_{\mathrm{ARK}}$
& PL$_{\mathrm{RNP}}$ \tabularnewline
\midrule

Pooled (100 samples)
& 60.59 
& 63.49
& 78.10
& 31.81
& 46.38
& 0.797 
& 59.67
& 24.44 \tabularnewline

\midrule

Protenix + \method
& 61.76
& 65.97
& 77.74
& 36.36
& 56.52
& 0.826
& 56.08
& 25.56 \tabularnewline

ESMFold2 + \method
& 62.94
& 67.78
& 78.10
& 39.77
& 50.72
& 0.818
& 60.53
& 28.89 \tabularnewline

OpenDDE + \method
& 68.24
& 63.69
& 78.47
& 27.27
& 49.28
& 0.720
& 64.35
& 24.44 \tabularnewline

\bottomrule
\end{tabular}
\vspace{-.1in}
\end{table*}

\paragraph{Exploiting representation complementarity for peer selection.}
Based on the representation complementarity observed above, we design peer selection to favor stronger complementary signals. Rather than using all peers with equal weights, we consider weighting peers by their interface representation magnitude and selecting peers with the lowest CKA to the anchor representation. In \Cref{tab:metric_ablation}, both strategies improve overall performance. However, they can also reduce performance on high-quality predictions for protein-protein complexes, as excluding peers may remove useful information and magnitude weighting may amplify incorrect signals. This suggests a trade-off in peer selection that depends on the task.

\subsection{Sample-level and seed-level scaling vs. \method}
\label{subsec:inference_scaling}

\begin{wrapfigure}{r}{0.35\textwidth}
\setcounter{figure}{9}
\vspace{-0.2in}
\centering
\includegraphics[width=0.94\linewidth]{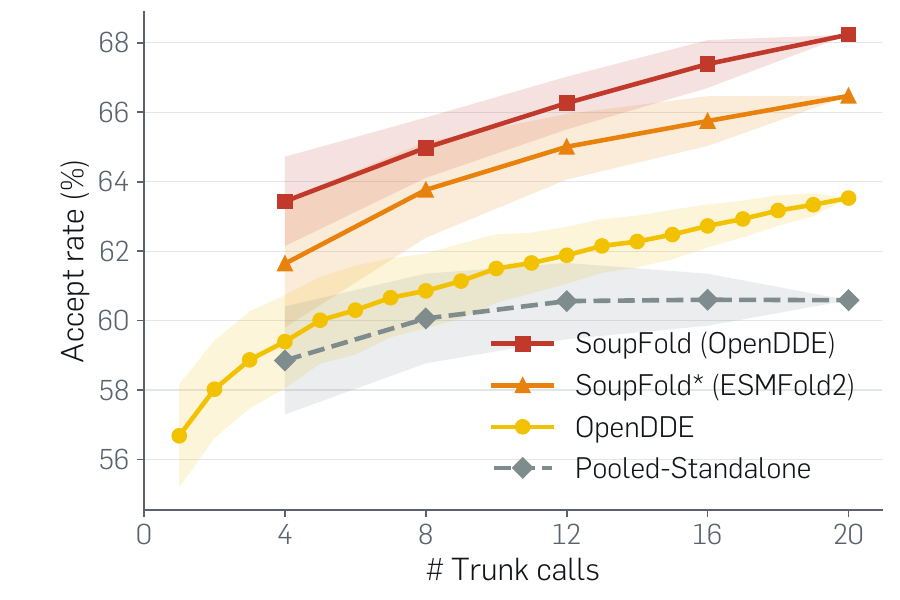}
\vspace{-.1in}
\caption{{\textbf{Scaling budget.}
Our method beats scaling OpenDDE.}}
\label{fig:scaling}
\vspace{-0.2in}
\end{wrapfigure}

We compare \method{} with sample-level ensembling across co-folding models, which pools predictions across multiple standalone models and selects the highest-confidence structure. As shown in \Cref{tab:pooled_samples}, despite comparing only 25 \method{} samples against 100 pooled standalone samples, \method{} is competitive or substantially better. This also reflects the miscalibration of confidence scores across models to select best-confidence samples. In contrast, \method{} combines information before structure generation and selects samples within a single anchor model.

We next compare \method{} with scaling a standalone model under a larger sampling budget. This is particularly important for antibody-antigen prediction, where prior work reports improved performance with more seeds \citep{esmfold2,abramson2024af3}. Each \method{} seed evaluates all four model trunks once, so we allocate the standalone model four times as many seeds, matching the number of trunk evaluations while giving it a larger diffusion sampling budget. We use OpenDDE, the strongest standalone model for antibody-antigen prediction, despite its substantially larger and more expensive trunk. As shown in \Cref{fig:scaling}, \method{} still outperforms scaling OpenDDE with the larger sampling budget.

\section{Related works}
\label{sec:background}

\paragraph{Protein co-folding models.} Modern co-folding models predict the structure of biomolecular complexes containing proteins, nucleic acids, and small molecules. AlphaFold3 \citep{abramson2024af3} introduce a co-folding framework for biomolecular structure prediction, followed by several models including, Boltz~\citep{wohlwend2024boltz,passaro2025boltz2}, Chai \citep{chai2024chai1}, Protenix~\citep{bytedance2025protenix}, OpenFold3~\citep{openfold3}, ESMFold2~\citep{esmfold2}, and OpenDDE~\citep{project2026foldingreasoningscalingopensource}. While Protenix closely follows the AlphaFold3 training recipe, ESMFold2 builds on representations from the pretrained 6B ESMC protein language model, and OpenDDE uses a multi-stage curriculum with emphasis on challenging interactions such as antibody-antigen complexes. 

\noindent \textbf{Representations in co-folding models.} Several studies have examined the internal representations of co-folding models. AlphaLink and AlphaLink2 \citep{stahl2023alphalink,stahl2024modelling} inject crosslink-derived distance restraints into pair representations to guide monomer and complex structure prediction toward specific conformations. More recently, \citet{jang2026systematic} use representations from Boltz as atom-level molecular features and show that they transfer to downstream tasks. \citet{suzuki2026pairscaling} show that scaling the pair representations of AlphaFold3 alters its conformational sampling. In contrast, we transfer representations across multiple co-folding models.

\noindent \textbf{Representation distillation and ensembling.}
Representation distillation transfers information between models by aligning intermediate representations of pretrained teacher and student models~\citep{vid}. Recent work has extended this idea to generative models for images \citep{yu2025repa} and protein structures \citep{kim2026triprorep}. Beyond distillation, independently trained models can be combined by averaging their parameters, as in Model Soups~\citep{wortsman2022model}, or by learning mappings between their representation spaces \citep{bansal2021revisiting}. %Our work builds on the latter perspective by mapping and combining representations from multiple co-folding models.

\section{Conclusion}
\label{sec:conclusion}

We study whether co-folding models encode complementary information in their internal representations and whether it can be transferred across models to improve structure prediction. We introduce \method{}, which learns lightweight mappings between pair-representation spaces and combines representations before structure generation. Without retraining the underlying models, \method{} improves prediction performance across standard, challenging, and higher-order complex benchmarks. Future work could explore adaptive peer weighting using representation-level interface uncertainty, or mixture-of-experts-style co-folding models specialized in complementary structural patterns.

%\section*{AI use statement}

%We used Claude Code to assist with author-specified data processing, implementation of author-specified methods, debugging, and resolving package conflicts or kernel issues when running co-folding models. We also used it to assist with visualizations and to polish figures and tables. It was not used to determine the methodology or experimental settings, which were determined by the authors. All AI-assisted outputs were reviewed and verified by the authors, and all results were interpreted by the authors. ChatGPT was used for writing and language polishing. The authors take full responsibility for the final content of this work.

%\section*{Ethics statement}

%This work could be misused to support the design of harmful compounds. All co-folding models used in this study were used for non-commercial academic research in accordance with their respective licenses or terms of use. This work uses publicly available biomolecular structure data and benchmarks and does not involve human participants, private data, or animal experiments.

%\section*{Reproducibility statement}

%We provide full details of the method, training, experimental settings, and evaluation procedures in the Appendix. The Supplementary Material includes the code and trained transfer-network weights.

\section*{Acknowledgments}

This work was supported by the AMD University Program’s AI \& HPC Program. This work was also supported by the Artificial Intelligence Specialized Foundation Model Project of the National IT Industry Promotion Agency (NIPA), funded by the Ministry of Science and ICT (MSIT) of the Republic of Korea (No. PJT-25-100009), and by the Institute for Information \& Communications Technology Planning \& Evaluation (IITP) grant funded by the Korea government (MSIT) (RS-2019-II190075, Artificial Intelligence Graduate School Program (KAIST)).

%\begin{abstract}
%\input{0_abstract}
%\end{abstract}

\bibliographystyle{plainnat}
\bibliography{resources/citation}

\newpage
\appendix
\section{Details of \method{}}
\label{sec:soupfold_appendix}\label{subsec:training_details}

%\input{figures/r2_score}

%\subsection{Training Transfer Networks}\label{subsec:training_details}

\paragraph{Training transfer networks.}
We train a transfer network $f_{m\rightarrow m^\star}$ for each directed pair of ESMFold2, AlphaFold3, Protenix, and OpenDDE, where $m$ denotes the peer model and $m^\star$ denotes the anchor model. Each map is a one-layer MLP of width $1024$, applied independently to each pair entry. The corresponding pair representation dimensions are $256$, $128$, $128$, and $384$, respectively.

Pair representations are obtained after ten recycling steps and standardized per channel using the channel-wise mean and standard deviation computed on the held-out training set. For model $m$, $H^{(m)}_{ij}\in\mathbb{R}^{C_m}$ denotes the pair representation for token pair $(i,j)$. Each transfer network is trained with
\begin{equation*}
\mathcal{L}(f_{m\rightarrow m^\star})
=
\mathbb{E}_{(H^{(m)},H^{(m^\star)})\in\mathcal{D}}
\left[
\mathbb{E}_{(i,j)}
\left\|
f_{m\rightarrow m^\star}(\hat H^{(m)}_{ij})
-
\hat H^{(m^\star)}_{ij}
\right\|_2^2
\right].
\end{equation*}
Here, $\hat H^{(m)}$ denotes the channel-wise standardized representation. Because the models use different tokenizations, we construct token maps across models and define the transfer loss only on corresponding pair entries. We use random crops of $512$ tokens with AdamW, a learning rate of $3\times10^{-4}$, no weight decay, and gradient clipping at $1.0$.

\textbf{Held-out training data for transfer networks.} We train the transfer maps on a held-out set collected from the RCSB PDB with deposition dates before September 30, 2021. We restrict the set to protein-only complexes with at least two chains. To reduce redundancy, we cluster distinct protein sequences using MMseqs2 at 40\% sequence identity and 80\% coverage. We then group complexes according to the sequence clusters of their constituent chains and retain one representative per complex family. From the resulting candidates, we select 3,000 structurally diverse complexes. 

For each complex, we prepare MSAs and template features using the same preprocessing pipeline across models. Templates are restricted to structures released before the corresponding cutoff. We extract pair representations from all four co-folding models after ten recycling iterations. The remaining systems are discarded so that every transfer map is trained on the same set of complexes. We train the transfer networks for five epochs.

\paragraph{Token maps.}
The four models may occasionally tokenize and order the same complex differently. Thus, we construct token maps between models using normalized token fingerprints based on chain, residue index, and representative atom. For each anchor-peer pair, the token map matches corresponding tokens and leaves unmatched or ambiguous tokens undefined. A pair entry $(i,j)$ is matched only when both tokens have valid correspondences.

\paragraph{Inference-time representation ensembling.}
At inference, we choose one model $m^\star$ as the anchor and use the remaining models as peers. Each model independently produces its pair representation $H^{(m)}$ after ten recycling iterations. Peer representations are aligned to the anchor token order using the token maps, standardized using their channel statistics, and mapped into the anchor representation space using the corresponding transfer networks. We then compute
\begin{equation*}
\hat H
=
\frac{
w_{m^\star}\hat H^{(m^\star)}
+
\sum_{m\neq m^\star}
w_m
f_{m\rightarrow m^\star}\!\left(\tilde{\hat H}^{(m)}\right)
}{
w_{m^\star}
+
\sum_{m\neq m^\star} w_m
},
\qquad
H'
=
\sigma^{(m^\star)}\odot\hat H
+
\mu^{(m^\star)},
\end{equation*}
where $\mu^{(m^\star)}$ and $\sigma^{(m^\star)}$ are the anchor channel statistics. Unmatched peers are excluded from both the numerator and denominator for the corresponding pair entry. We use uniform weights $w_{m^\star}=w_m=1$ throughout. The resulting $H'$ replaces the anchor pair representation, while its single representation and all other inputs remain unchanged. The anchor diffusion module then generates the final structure conditioned on $H'$.

\newpage

\section{Model Configuration}\label{appx:configuration}

\subsection{Input preparation}

\paragraph{Sequences.}
We construct all inputs from the official FoldBench ground-truth mmCIF. We first construct one canonical record. From this record, we generate the model-specific input for each of the four models. Thus, all four models receive the same chains, sequences, and ligands. Polymer sequences are obtained from SEQRES. Ligands, ions, and cofactors are included with their deposited copy numbers. Non-standard polymer residues are represented as modifications, except residues canonicalized to their parent amino acid by the reference pipelines (MSE, SEC, and PYL).

\paragraph{MSAs.}
We generate multiple sequence alignments once using the ColabFold MMseqs2 server with UniRef30 and environmental databases and share them across all models. For complexes containing at least two protein chains, including homo-oligomers, we additionally generate a paired A3M. Following the Protenix convention, paired alignments are also appended to the unpaired MSA block using \texttt{msa\_pair\_as\_unpair = True}.

\paragraph{Templates.}
Template search follows the AlphaFold~3 pipeline.
We run \texttt{hmmbuild} and \texttt{hmmsearch} with the AF3 settings against the PDB SeqRes database, with one search per unique sequence.
We retain only PDB entries released on or before September 30, 2021.
The top template hits are retrieved as mmCIF files from PDBe and stored locally.
Each model then applies its own template filtering and uses at most four templates per chain.

\subsection{Models, checkpoints, and source modifications}

\begin{table}[h]
\centering
\small
\begin{tabular}{llcl}
\toprule
Model & Checkpoint & $C_z$  \\
\midrule
AlphaFold~3 & alphafold3 3.0.1, released weights & 128 \\
OpenDDE & \texttt{opendde\_v1} & 384 \\
Protenix & \texttt{protenix\_base\_default\_v1.0.0} & 128  \\
ESMFold2 & esm 3.4.0, ESMFold2 + ESMC-6B language model & 256 \\
\bottomrule
\end{tabular}
\caption{Packages, checkpoints, and pair-representation width $C_z$.}
\label{tab:app-models}
\end{table}

\paragraph{Source modifications for consistent tokenization.}

For \method exclusively, we make minor modifications to three packages so that their pair representations are better aligned token-by-token. 
\begin{itemize}[topsep=-1.0pt,itemsep=1.0pt,leftmargin=3.5mm]
  \item \textbf{AF3} (\texttt{atom\_layout.py}, 2 lines): we retain the glycan O1 atom, which AF3 otherwise removes and which changes the token count for glycan-containing systems.
  \item \textbf{Protenix} and \textbf{OpenDDE} (\texttt{data/core/parser.py}, 8 and 4 lines): entities specified as ligands or ions are tokenized per atom even when their CCD type is ``protein''. This handles free amino-acid ligands such as ILE, VAL, and TRP consistently with AF3 and ESMFold2.
  \item \textbf{ESMFold2}: no source modifications.
\end{itemize}

\subsection{Inference Configuration}

By default, we use $10$ recycling steps. We use the default number of diffusion steps for each co-folding model. We generate $25$ samples per target. We select the highest-confidence prediction using each model's default confidence score. Each system produces five diffusion samples per seed.

\subsection{Representation extraction}

For each system and model, we run the full network once with 10 recycling iterations and store the final trunk pair representation $z \in \mathbb{R}^{N \times N \times C_z}$. All representations are stored in fp32.

\section{Benchmark}\label{appx:benchmark}

\paragraph{FoldBench.}

We use the official FoldBench protocol \citep{foldbench2025}. We evaluate the protein-protein, antibody-antigen, and protein-ligand subsets. Ligands are specified by their CCD codes. We report results on the systems for which all four co-folding models produce a prediction: $274$ interfaces for protein-protein, $170$ interfaces for antibody-antigen, and $526$ poses for protein-ligand complexes. Following official metrics, for protein-protein and antibody-antigen complexes, we compute DockQ with \texttt{DockQ v2.1.3}~\citep{mirabello2024dockq} and report the success rate at DockQ $\geq 0.23$, $0.49$, and $0.80$. For protein-ligand complexes, we compute lDDT-PLI and LRMSD (BiSyRMSD) with OpenStructure and count a ligand as successful when LRMSD $< 2$\,\AA{} and lDDT-PLI $> 0.8$.

\paragraph{MGBench.}

We evaluate molecular glue-induced ternary complexes using MGBench~\citep{liao2025benchmarking}. We use the published ternary protein chains and molecular-glue annotations and evaluate the $88$ complexes successfully processed by all four co-folding models. Each system contains the two annotated protein chains and the molecular glue specified by its CCD code, and the deposited mmCIF structure is used as ground truth. Following official metrics, we evaluate the protein-protein interface using DockQ and define protein-protein success as DockQ $\geq 0.23$. For the molecular-glue pose, we compute LRMSD and lDDT-PLI with OpenStructure, and define protein-ligand success as LRMSD $<2$~\AA{} and lDDT-PLI $>0.8$. We additionally consider ternary success that requires both protein-protein and protein-ligand success for the same predicted structure.

\paragraph{GPCR.}

We evaluate GPCR complexes using the benchmark from the recent study \citep{shen2025update}. We consider the $69$ systems successfully processed by all four co-folding models. Following benchmark settings, each input contains the receptor, the transducer when present, and the bound small molecules specified by their CCD codes. Following \citet{shen2025update}, crystallization aids and receptor fusion proteins are removed, with fusion regions replaced by the native receptor sequence. Following the official measures, we use TM6 RMSD and ligand LRMSD. We align predictions to the ground truth using the C$\alpha$ atoms of TM1-5 and TM7 and compute the TM6 C$\alpha$ RMSD. We additionally define P-L success as LRMSD $<2$~\AA{} for every ligand in the complex. We additionally define TM6 success as a C$\alpha$ RMSD $<3$~\AA{}. We additionally consider TM6-P-L success that requires both criteria to be satisfied by the same prediction.

\paragraph{CASP16 oligomer.}

We use the CASP16 \citep{zhang2026assessment} oligomer targets with their published stoichiometry and the official target structures as ground truth. We evaluate the $18$ targets for which all four co-folding models produce a prediction. Following official metrics, we score each prediction with OpenStructure \texttt{compare-structures} and report the CASP16 assembly metrics: interface DockQ weighted by the number of native contacts, ICS, IPS, QS-best, lDDT, and TM-score.

\paragraph{ARK-AB.}

We use the ARK-AB antibody-antigen benchmark released with OpenDDE~\citep{project2026foldingreasoningscalingopensource}, with its interface definitions. Each input is the first biological assembly of the entry. We evaluate the $162$ entries for which all four co-folding models produce a prediction, comprising $398$ interfaces in $155$ interface clusters. Interface clusters are defined by the benchmark as unordered pairs of MMseqs2 entity clusters ($40\,\%$ sequence identity, $80\,\%$ coverage). We compute DockQ for each interface and report the success rate at DockQ $\geq 0.23$, $0.49$, and $0.80$. Following the official metrics, the success rate is normalized over interface clusters: we compute the fraction of successful interfaces within each cluster and average this fraction over clusters.

\paragraph{Runs N' Poses.}

We use the challenging subset, $0$-$20$\,\% similarity bin, of Runs N' Poses benchmark~\citep{vskrinjar2026evaluating}. Here, ligands are specified by their CCD codes. We evaluate the $83$ systems ($90$ ligands) for which all four co-folding models produce a prediction. Following official metrics, we compute lDDT-PLI and BiSyRMSD with OpenStructure and count a ligand as successful when BiSyRMSD $< 2$\,\AA{} and lDDT-PLI $> 0.8$.

\newpage

\section{Additional Results}\label{appx:additional_results}

\subsection{MGBench}\label{appx:mgbench_results}

% ============================================================
% Table: MGBench
% ============================================================

\begin{table*}[h]
\centering
\small
\setlength{\tabcolsep}{6.0pt}

\caption{%
\textbf{Performance on MGBench molecular-glue ternary complexes.}
We evaluate the $88$ molecular glue ternary complexes for which all four co-folding models produce a prediction.
PLP counts a system as successful only when PP (DockQ $\geq 0.23$) and PL (LRMSD $< 2$\,\AA{} and lDDT-PLI $> 0.8$) hold for the same structure.
The first line in each cell reports the highest-confidence prediction among 25 samples (5 samples per seed, 5 seeds).
The second line reports the mean $\pm$ standard deviation over the 25 samples.
}

\label{tab:mgbench_full}

\begin{tabular}{lcccccc}
\toprule
Model & PLP (\%) & PP (\%) & PL (\%) & DockQ $\uparrow$ & LRMSD $\downarrow$ & lDDT-PLI $\uparrow$ \\
\midrule

AlphaFold3
& \mainnewscore{31.82}{33.91}{1.05}
& \mainnewscore{47.73}{49.64}{2.63}
& \mainnewscore{32.95}{34.05}{0.99}
& \shortstack[c]{0.439\\{\scriptsize $0.452 \pm 0.011$}}
& \shortstack[c]{4.77\\{\scriptsize $5.10 \pm 0.69$}}
& \shortstack[c]{0.637\\{\scriptsize $0.645 \pm 0.010$}}
\\

\midrule

ESMFold2
& \mainnewscore{30.68}{26.55}{2.13}
& \mainnewscore{\textbf{53.41}}{48.59}{1.34}
& \mainnewscore{31.82}{26.59}{2.11}
& \shortstack[c]{0.481\\{\scriptsize $0.453 \pm 0.009$}}
& \shortstack[c]{4.94\\{\scriptsize $5.71 \pm 0.70$}}
& \shortstack[c]{0.669\\{\scriptsize $0.659 \pm 0.009$}}
\\

\rowcolor{soupgray}
\methodrow
& \mainnewscore{\textbf{39.77}}{34.14}{2.66}
& \mainnewscore{52.27}{51.36}{1.64}
& \mainnewscore{\textbf{39.77}}{34.18}{2.73}
& \shortstack[c]{\textbf{0.487}\\{\scriptsize $0.466 \pm 0.010$}}
& \shortstack[c]{\textbf{4.38}\\{\scriptsize $4.98 \pm 0.82$}}
& \shortstack[c]{\textbf{0.702}\\{\scriptsize $0.664 \pm 0.016$}}
\\

\midrule

Protenix
& \mainnewscore{28.41}{31.18}{2.34}
& \mainnewscore{50.00}{49.50}{1.25}
& \mainnewscore{28.41}{31.45}{2.40}
& \shortstack[c]{0.459\\{\scriptsize $0.454 \pm 0.008$}}
& \shortstack[c]{\textbf{5.76}\\{\scriptsize $5.48 \pm 0.23$}}
& \shortstack[c]{0.656\\{\scriptsize $0.663 \pm 0.017$}}
\\

\rowcolor{soupgray}
\methodrow
& \mainnewscore{\textbf{36.36}}{35.86}{1.16}
& \mainnewscore{\textbf{54.55}}{51.91}{1.27}
& \mainnewscore{\textbf{36.36}}{36.05}{1.14}
& \shortstack[c]{\textbf{0.479}\\{\scriptsize $0.475 \pm 0.007$}}
& \shortstack[c]{5.93\\{\scriptsize $4.96 \pm 0.35$}}
& \shortstack[c]{\textbf{0.672}\\{\scriptsize $0.683 \pm 0.007$}}
\\

\midrule

OpenDDE
& \mainnewscore{\textbf{31.82}}{31.09}{1.54}
& \mainnewscore{50.00}{57.55}{3.08}
& \mainnewscore{\textbf{31.82}}{31.09}{1.54}
& \shortstack[c]{0.471\\{\scriptsize $0.491 \pm 0.013$}}
& \shortstack[c]{5.45\\{\scriptsize $4.96 \pm 0.27$}}
& \shortstack[c]{\textbf{0.685}\\{\scriptsize $0.694 \pm 0.009$}}
\\

\rowcolor{soupgray}
\methodrow
& \mainnewscore{27.27}{27.41}{1.86}
& \mainnewscore{\textbf{52.27}}{55.23}{2.41}
& \mainnewscore{27.27}{27.41}{1.86}
& \shortstack[c]{\textbf{0.484}\\{\scriptsize $0.489 \pm 0.010$}}
& \shortstack[c]{\textbf{4.93}\\{\scriptsize $4.90 \pm 0.18$}}
& \shortstack[c]{0.683\\{\scriptsize $0.688 \pm 0.005$}}
\\

\bottomrule
\end{tabular}

\end{table*}

\newpage

\subsection{GPCR}\label{appx:gpcr_results}

\begin{table}[h]
\centering
\setlength{\tabcolsep}{3.5pt}
\caption{%
\textbf{Performance on the GPCR benchmark.}
The benchmark comprises $69$ complexes for which all four co-folding models produce a prediction.
TM6 and TM6 tip are the mean C$\alpha$ RMSD (\AA) of the TM6 helix and of its cytoplasmic end, after superposition on TM1-5 and TM7.
Following official metrics, P-L success requires LRMSD $<2$~\AA\ for every ligand in the complex. 
TM6-P-L additionally requires TM6-tip RMSD $<3.0$~\AA\ in the same prediction.
We report the best-confidence result (5 samples per seed, 5 seeds), and the mean $\pm$ standard deviation over the 25 samples.
}
\label{tab:gpcr_full}
\begin{tabular}{l ccccc}
\toprule
Model
& TM6-P-L$\uparrow$ & TM6~(\AA)$\downarrow$ & TM6 tip~(\AA)$\downarrow$ & P-L$\uparrow$ & LRMSD~(\AA)$\downarrow$
\tabularnewline
\midrule
AlphaFold3
& \mainnewscore{49.28}{48.46}{2.60}
& \mainnewscore{1.24}{1.22}{.01}
& \mainnewscore{1.39}{1.38}{.02}
& \mainnewscore{53.62}{50.90}{2.29}
& \mainnewscore{5.47}{5.06}{.53}
\tabularnewline
\midrule
Protenix
& \mainnewscore{49.28}{50.55}{2.73}
& \mainnewscore{1.32}{1.34}{.04}
& \mainnewscore{1.58}{1.62}{.07}
& \mainnewscore{50.72}{53.68}{2.54}
& \mainnewscore{6.13}{5.95}{.61}
\tabularnewline
\rowcolor{soupgray}
+\method
& \mainnewscoregain{\textbf{56.52}}{51.01}{3.28}{7.25}
& \mainnewscoregain{\textbf{1.19}}{1.20}{.02}{.13}
& \mainnewscoregain{\textbf{1.42}}{1.44}{.03}{.16}
& \mainnewscoregain{\textbf{57.97}}{53.28}{2.91}{7.25}
& \mainnewscoregain{\textbf{4.49}}{5.04}{.71}{1.64}
\tabularnewline
\midrule
ESMFold2
& \mainnewscore{43.48}{42.38}{2.57}
& \mainnewscore{1.33}{1.36}{.03}
& \mainnewscore{1.72}{1.73}{.07}
& \mainnewscore{47.83}{45.86}{2.24}
& \mainnewscore{5.33}{5.45}{.33}
\tabularnewline
\rowcolor{soupgray}
+\method
& \mainnewscoregain{\textbf{50.72}}{48.46}{2.13}{7.25}
& \mainnewscoregain{\textbf{1.19}}{1.19}{.02}{.15}
& \mainnewscoregain{\textbf{1.40}}{1.40}{.04}{.32}
& \mainnewscoregain{\textbf{53.62}}{50.20}{2.39}{5.80}
& \mainnewscoregain{\textbf{4.15}}{4.83}{.39}{1.19}
\tabularnewline
\midrule
OpenDDE
& \mainnewscore{46.38}{42.84}{2.78}
& \mainnewscore{1.39}{1.35}{.03}
& \mainnewscore{1.59}{1.56}{.05}
& \mainnewscore{47.83}{46.49}{2.20}
& \mainnewscore{5.75}{5.35}{.32}
\tabularnewline
\rowcolor{soupgray}
+\method
& \mainnewscoregain{\textbf{49.28}}{49.45}{2.50}{2.90}
& \mainnewscoregain{\textbf{1.18}}{1.21}{.03}{.20}
& \mainnewscoregain{\textbf{1.33}}{1.41}{.05}{.25}
& \mainnewscoregain{\textbf{50.72}}{51.30}{2.21}{2.90}
& \mainnewscoregain{\textbf{5.15}}{5.26}{.58}{.59}
\tabularnewline
\bottomrule
\end{tabular}
\end{table}

\newpage

\subsection{CASP16 Oligomer Benchmarks}\label{appx:casp16_results}

% ============================================================
% Table: CASP16 oligomer benchmark
% ============================================================
\begin{table*}[h]
\centering
\small
\setlength{\tabcolsep}{4.2pt}

\caption{%
\textbf{Performance on CASP16 oligomer complexes.}
The benchmark comprises $18$ oligomer complexes for which all four co-folding models produce a prediction.
The first line in each cell reports the highest-confidence prediction among 25 samples (5 samples per seed, 5 seeds). The second line reports the mean $\pm$ standard deviation over the 25 samples. Following official metrics, DockQ aggregates interface-level DockQ across each oligomer, weighting interface $i$ by $\log_{10}(\mathrm{\text{number of interacting residues}_i}/2)$.
}
\label{tab:casp16_oligomer}

\begin{tabular}{l cccccc}
\toprule
Model & DockQ & lDDT & TM-score & QSbest & ICS & IPS
\tabularnewline
\midrule
AlphaFold3
& \mainnewscore{0.5106}{0.4783}{0.0226}
& \mainnewscore{0.8326}{0.8306}{0.0027}
& \mainnewscore{{0.8379}}{0.8016}{0.0171}
& \mainnewscore{0.6375}{0.6424}{0.0157}
& \mainnewscore{0.5116}{0.5126}{0.0097}
& \mainnewscore{0.5983}{0.5882}{0.0073}
\tabularnewline
\midrule
ESMFold2
& \mainnewscore{0.4500}{0.4052}{0.0275}
& \mainnewscore{0.7830}{0.7644}{0.0073}
& \mainnewscore{0.7521}{0.6925}{0.0329}
& \mainnewscore{0.6051}{0.5703}{0.0314}
& \mainnewscore{0.5024}{0.4751}{0.0222}
& \mainnewscore{0.5606}{0.5489}{0.0188}
\tabularnewline

\rowcolor{soupgray}
\methodrow
& \mainnewscore{\textbf{0.4890}}{0.4457}{0.0201}%{0.0390}
& \mainnewscore{\textbf{0.8218}}{0.7935}{0.0127}%{0.0388}
& \mainnewscore{\textbf{0.8182}}{0.7566}{0.0255}%{0.0661}
& \mainnewscore{\textbf{0.6592}}{0.6028}{0.0308}%{0.0541}
& \mainnewscore{\textbf{0.5117}}{0.4754}{0.0192}%{0.0093}
& \mainnewscore{\textbf{0.5891}}{0.5648}{0.0114}%{0.0285}
\tabularnewline

\midrule

Protenix
& \mainnewscore{0.4883}{0.4426}{0.0195}
& \mainnewscore{0.8155}{0.8066}{0.0035}
& \mainnewscore{0.8261}{0.7600}{0.0263}
& \mainnewscore{0.6553}{0.6160}{0.0243}
& \mainnewscore{0.5151}{0.4887}{0.0150}
& \mainnewscore{0.5681}{0.5523}{0.0120}
\tabularnewline

\rowcolor{soupgray}
\methodrow
& \mainnewscore{\textbf{0.5512}}{0.4847}{0.0258}%{0.0629}
& \mainnewscore{\textbf{0.8353}}{0.8266}{0.0040}%{0.0198}
& \mainnewscore{\textbf{0.8324}}{0.8148}{0.0257}%{0.0063}
& \mainnewscore{\textbf{0.6971}}{0.6568}{0.0253}%{0.0418}
& \mainnewscore{\textbf{0.5497}}{0.5201}{0.0170}%{0.0346}
& \mainnewscore{\textbf{0.6089}}{0.5970}{0.0144}%{0.0408}
\tabularnewline

\midrule

OpenDDE
& \mainnewscore{0.4537}{0.4468}{0.0120}
& \mainnewscore{0.6426}{0.6438}{0.0024}
& \mainnewscore{0.6507}{0.6581}{0.0228}
& \mainnewscore{0.5062}{0.4992}{0.0134}
& \mainnewscore{0.4109}{0.4071}{0.0095}
& \mainnewscore{0.5114}{0.5102}{0.0109}
\tabularnewline

\rowcolor{soupgray}
\methodrow
& \mainnewscore{\textbf{0.5304}}{0.4750}{0.0202}%{0.0767}
& \mainnewscore{\textbf{0.6673}}{0.6585}{0.0024}%{0.0247}
& \mainnewscore{\textbf{0.7207}}{0.6873}{0.0184}%%{0.0700}
& \mainnewscore{\textbf{0.5688}}{0.5288}{0.0187}%{0.0626}
& \mainnewscore{\textbf{0.4647}}{0.4316}{0.0169}%{0.0538}
& \mainnewscore{\textbf{0.5362}}{0.5252}{0.0106}%{0.0248}
\tabularnewline

\bottomrule
\end{tabular}
\end{table*}

\newpage

\subsection{ARK Antibody-Antigen Benchmarks}\label{appx:ark_results}

\begin{table*}[h]
\centering
\small
\setlength{\tabcolsep}{5.2pt}
\small
\caption{%
\textbf{Performance on ARK-AB benchmark.}
The benchmark comprises $162$ recent, low-homology PDB complexes, decomposed into $398$ antibody-antigen interfaces grouped into $155$ interface clusters for which all four co-folding models produce a prediction.
We report success at three DockQ quality tiers: acceptable ($\ge 0.23$), medium ($\ge 0.49$), and high ($\ge 0.80$).
\textit{Cluster success (default)} averages the per-cluster success rate over the clusters, while \textit{Interface success} reports the raw success rate over all interfaces.
For each metric, we report the best-confidence result (5 samples per seed, 5 seeds) and the mean $\pm$ standard deviation over the 25 samples.
}
\label{tab:soupfold_arkab}

\begin{tabular}{l cc cc}
\toprule
& \multicolumn{2}{c}{Cluster Success}
& \multicolumn{2}{c}{Interface Success}
\tabularnewline
\cmidrule(lr){2-3}
\cmidrule(lr){4-5}
Model
& Best-conf. & Mean $\pm$ Std
& Best-conf. & Mean $\pm$ Std
\tabularnewline
\midrule
\multicolumn{5}{l}{\textit{DockQ} $>0.23$ (acceptable)}
\tabularnewline
\midrule
AlphaFold3
& 47.66 & 42.27 $\pm$ 2.00
& 58.04 & 51.80 $\pm$ 2.05
\tabularnewline
Protenix
& 45.94 & 42.08 $\pm$ 1.66
& 57.54 & 53.29 $\pm$ 1.23
\tabularnewline
\rowcolor{soupgray}
+\method
& \textbf{56.08} & 54.02 $\pm$ 3.37
& \textbf{67.34} & 64.17 $\pm$ 2.86
\tabularnewline
ESMFold2
& 48.68 & 39.91 $\pm$ 2.53
& 59.30 & 47.70 $\pm$ 2.17
\tabularnewline
\rowcolor{soupgray}
+\method
& \textbf{60.53} & 52.38 $\pm$ 2.08
& \textbf{69.85} & 61.31 $\pm$ 1.58
\tabularnewline
OpenDDE
& 58.92 & 57.51 $\pm$ 1.60
& 68.59 & 64.91 $\pm$ 1.56
\tabularnewline
\rowcolor{soupgray}
+\method
& \textbf{64.35} & 61.37 $\pm$ 1.75
& \textbf{73.12} & 70.24 $\pm$ 1.60
\tabularnewline
\midrule
\multicolumn{5}{l}{\textit{DockQ} $>0.49$ (medium)}
\tabularnewline
\midrule
AlphaFold3
& 41.62 & 36.29 $\pm$ 1.72
& 53.77 & 47.67 $\pm$ 1.64
\tabularnewline
Protenix
& 37.85 & 34.03 $\pm$ 1.40
& 50.75 & 46.61 $\pm$ 1.18
\tabularnewline
\rowcolor{soupgray}
+\method
& \textbf{47.48} & 46.60 $\pm$ 2.38
& \textbf{60.80} & 58.31 $\pm$ 2.18
\tabularnewline
ESMFold2
& 42.88 & 34.66 $\pm$ 2.27
& 54.52 & 43.24 $\pm$ 2.12
\tabularnewline
\rowcolor{soupgray}
+\method
& \textbf{53.54} & 44.75 $\pm$ 1.91
& \textbf{64.07} & 55.25 $\pm$ 1.46
\tabularnewline
OpenDDE
& 55.73 & 53.84 $\pm$ 1.73
& 65.33 & 61.81 $\pm$ 1.42
\tabularnewline
\rowcolor{soupgray}
+\method
& \textbf{57.74} & 56.34 $\pm$ 2.09
& \textbf{67.84} & 65.23 $\pm$ 1.62
\tabularnewline
\midrule
\multicolumn{5}{l}{\textit{DockQ} $>0.80$ (high)}
\tabularnewline
\midrule
AlphaFold3
& 21.60 & 19.54 $\pm$ 1.49
& 34.92 & 30.58 $\pm$ 1.30
\tabularnewline
Protenix
& 19.97 & 16.42 $\pm$ 1.33
& 31.91 & 27.69 $\pm$ 1.20
\tabularnewline
\rowcolor{soupgray}
+\method
& \textbf{20.42} & 19.89 $\pm$ 1.38
& \textbf{34.92} & 32.89 $\pm$ 1.08
\tabularnewline
ESMFold2
& 19.40 & 13.42 $\pm$ 1.81
& 32.66 & 23.65 $\pm$ 1.90
\tabularnewline
\rowcolor{soupgray}
+\method
& \textbf{20.40} & 18.92 $\pm$ 1.64
& \textbf{34.67} & 32.14 $\pm$ 1.34
\tabularnewline
OpenDDE
& \textbf{32.18} & 30.15 $\pm$ 1.33
& \textbf{43.47} & 40.86 $\pm$ 1.09
\tabularnewline
\rowcolor{soupgray}
+\method
& 30.41 & 29.28 $\pm$ 1.14
& 43.22 & 41.53 $\pm$ 0.77
\tabularnewline
\bottomrule
\end{tabular}
\vspace{-.1in}
\end{table*}

\newpage

\subsection{Low-similarity Subset of Runs N' Poses}\label{appx:rnp_results}

% ============================================================
% Table: RNP
% ============================================================

\begin{table*}[h]
\centering
\small
\setlength{\tabcolsep}{6.0pt}

\caption{%
\textbf{Performance on low-similarity subset of Runs N' Poses.}
We evaluate the $[0,20)$ low-similarity subset of Runs N' Poses benchmarks. We use 90 ligands from 83 systems for which all four co-folding models produce a prediction.
The first line in each cell reports the highest-confidence prediction among 25 samples
(5 samples per seed, 5 seeds).
The second line reports the mean $\pm$ standard deviation over the 25 samples.
}

\label{tab:rnp_full}

\begin{tabular}{lccc}
\toprule
Model & Success (\%) & BiSyRMSD $\downarrow$ & lDDT-PLI $\uparrow$ \\
\midrule

AlphaFold3
& \mainnewscore{28.89}{25.64}{1.63}
& \shortstack[c]{11.27\\{\scriptsize $11.14 \pm 0.54$}}
& \shortstack[c]{0.40\\{\scriptsize $0.38 \pm 0.01$}}
\\

\midrule

ESMFold2
& \mainnewscore{27.78}{25.33}{1.47}
& \shortstack[c]{12.20\\{\scriptsize $12.31 \pm 0.37$}}
& \shortstack[c]{0.37\\{\scriptsize $0.35 \pm 0.01$}}
\\

\rowcolor{soupgray}
\methodrow
& \mainnewscore{\textbf{28.89}}{25.06}{1.97}
& \shortstack[c]{\textbf{10.91}\\{\scriptsize $11.29 \pm 0.48$}}
& \shortstack[c]{\textbf{0.41}\\{\scriptsize $0.39 \pm 0.01$}}
\\

\midrule

Protenix
& \mainnewscore{21.11}{21.78}{0.89}
& \shortstack[c]{11.67\\{\scriptsize $11.41 \pm 0.31$}}
& \shortstack[c]{0.38\\{\scriptsize $0.39 \pm 0.01$}}
\\

\rowcolor{soupgray}
\methodrow
& \mainnewscore{\textbf{25.56}}{24.93}{1.71}
& \shortstack[c]{\textbf{11.25}\\{\scriptsize $11.62 \pm 0.53$}}
& \shortstack[c]{\textbf{0.40}\\{\scriptsize $0.39 \pm 0.01$}}
\\

\midrule

OpenDDE
& \mainnewscore{15.56}{18.02}{2.22}
& \shortstack[c]{10.42\\{\scriptsize $10.17 \pm 0.48$}}
& \shortstack[c]{0.35\\{\scriptsize $0.36 \pm 0.02$}}
\\

\rowcolor{soupgray}
\methodrow
& \mainnewscore{\textbf{24.44}}{24.58}{1.59}
& \shortstack[c]{\textbf{9.86}\\{\scriptsize $\mathbf{10.13} \pm 0.41$}}
& \shortstack[c]{\textbf{0.39}\\{\scriptsize $\mathbf{0.39} \pm 0.01$}}
\\

\bottomrule
\end{tabular}

\end{table*}

\newpage

\subsection{Transfer Training}\label{appx:transfer}

% ============================================================
% Representation transfer map ablation
% ============================================================

\begin{table}[h]
\centering
\small
\setlength{\tabcolsep}{4.5pt}

\caption{%
\textbf{Training representation transfer maps.}
Ablation on FoldBench antibody-antigen and protein-protein complexes using ESMFold2 as the anchor model.
The default maps are trained once on a held-out set.
We compare them with maps trained directly on representations from each FoldBench benchmark.
Accept, Medium, and High correspond to DockQ $\geq 0.23$, $\geq 0.49$, and $\geq 0.80$, respectively.
The first line in each cell reports the best-confidence-of-25 result (5 samples per seed, 5 seeds), and the second line reports the mean $\pm$ standard deviation over the 25 samples.
}

\label{tab:fmap_ablation}

\begin{tabular}{ll ccc}
\toprule
& & \multicolumn{3}{c}{DockQ success rate (\%)} \tabularnewline
\cmidrule(lr){3-5}
Benchmark & Training data for $f$ & Accept & Medium & High \tabularnewline
\midrule

\multirow{2}{*}{Ab-Ag}
& \raisebox{1.2ex}[0pt][0pt]{Held-out (default)}
& \mainnewscore{\textbf{62.94}}{52.96}{2.47}
& \mainnewscore{\textbf{51.18}}{43.58}{1.71}
& \mainnewscore{\textbf{24.12}}{20.12}{1.70}
\tabularnewline

& \raisebox{1.2ex}[0pt][0pt]{FoldBench Antibody-Antigen}
& \mainnewscore{61.18}{51.76}{2.54}
& \mainnewscore{48.82}{41.91}{1.68}
& \mainnewscore{23.53}{19.08}{1.97}
\tabularnewline

\midrule

\multirow{2}{*}{P-P}
& \raisebox{1.2ex}[0pt][0pt]{Held-out (default)}
& \mainnewscore{\textbf{78.10}}{75.33}{0.94}
& \mainnewscore{\textbf{72.63}}{70.22}{0.95}
& \mainnewscore{48.54}{45.07}{1.79}
\tabularnewline

& \raisebox{1.2ex}[0pt][0pt]{FoldBench Protein-Protein}
& \mainnewscore{\textbf{78.10}}{75.28}{1.00}
& \mainnewscore{\textbf{72.63}}{69.71}{0.88}
& \mainnewscore{\textbf{49.27}}{45.20}{1.47}
\tabularnewline

\bottomrule
\end{tabular}

\end{table}

%\newpage

%\section{Collapse and Recovery}\label{appx:collapse}

\end{document}